\documentclass[floats, eqnum, showpacs, nofootinbib, 
preprint,
eqsecnum ]{revtex4-1}

\usepackage{color,graphicx}
\usepackage{amsfonts}
\usepackage{amssymb}
\begin{document}

\title{
Light curves of light rays passing through a wormhole}

\author{Naoki Tsukamoto${}^{1,2}$}\email{tsukamoto@rikkyo.ac.jp}

\author{Tomohiro Harada${}^{2}$}\email{harada@rikkyo.ac.jp}

\affiliation{
${}^{1}$School of Physics, Huazhong University of Science and Technology, Wuhan 430074, China\\
${}^{2}$Department of Physics, Rikkyo University, Tokyo 171-8501, Japan
}
\date{\today}

\begin{abstract}
Gravitational lensing  is a good probe into the topological structure of dark gravitating celestial objects. 
In this paper, we investigate the light curve of a light ray that passes through the throat of an Ellis wormhole, the simplest example of traversable wormholes. 
The method developed here is also applicable to other traversable wormholes. 
To study whether the light curve of a light ray that passes through a wormhole throat is distinguishable from that which does not, 
we also calculate light curves without the passage of a throat  for an Ellis wormhole, a Schwarzschild black hole, and an ultrastatic wormhole with the spatial 
geometry identical to that of the Schwarzschild black hole in the  following two cases: 
(i) ``microlensing,'' where the  source, lens, and observer are almost aligned in this order and 
the light ray starts at the source, 
refracts in the weak gravitational field of the lens with a small 
 deflection angle, and reaches the observer; 
 and (ii) ``retrolensing,'' where the source, observer, and lens are 
almost aligned in this order, and the light ray starts at the source,
 refracts in the vicinity of the light sphere of the 
 lens with a deflection angle very close to $\pi$, and reaches the observer. 
We find that the light curve of the light ray that passes through the throat of the 
Ellis wormhole is clearly distinguishable from that by the 
microlensing but not from that by the retrolensing. 
This is because the light curve 
of a light ray that passes by a light sphere of a lens
with a large deflection angle has common characters, 
irrespective of the details of the lens object.
This implies that the light curves of the light rays that pass
through the throat of more general traversable wormholes  
are qualitatively the same as that of the Ellis wormhole.
\end{abstract}


\preprint{
RUP-16-21
}

\maketitle

\section{Introduction}
It is well known that general relativity admits the nontrivial topology of spacetimes like wormhole spacetimes
as nonvacuum solutions of the Einstein equation. 
(See Visser \cite{Visser_1995} for the details of wormholes.) 
The investigation of observational methods to find wormholes is important to understand our Universe.
We can survey wormholes with their gravitational lensing effects. 
(See~\cite{Schneider_Ehlers_Falco_1992,Petters_Levine_Wambsganss_2001,Perlick_2004_Living_Rev,Schneider_Kochanek_Wambsganss_2006,Bartelmann_2010} 
and references therein for the details of the gravitational lens.)
Kim and Cho~\cite{Kim_Cho_1994} and Cramer \textit{et al.}~\cite{Cramer_Forward_Morris_Visser_Benford_Landis_1995} 
pioneered the gravitational lensing effects of wormholes.
The gravitational lenses of wormholes with a negative mass ~\cite{Cramer_Forward_Morris_Visser_Benford_Landis_1995,
Safonova_Torres_Romero_2001_Jan,Takahashi_Asada_2013}
and with a positive mass~\cite{Rahaman_Kalam_Chakraborty_2007,Nandi_Zhang_Zakharov_2006,Dey_Sen_2008,Tejeiro_Larranaga_2012}
have been investigated for the last two decades.

The Ellis wormhole~\cite{Ellis_1973,Bronnikov_1973} 
is one of the simplest wormhole solutions of the Einstein equation 
with a ghost scalar field and belongs to the Morris-Thorne 
class~\cite{Morris_Thorne_1988,Morris_Thorne_Yurtsever_1988}.
This wormhole is shown to be unstable against spherical 
perturbations~\cite{Shinkai_Hayward_2002}.
On the other hand, 
some wormhole solutions have the metric that is identical to that of
the Ellis wormhole as their simplest
cases~\cite{Das:2005un,Shatskiy:2008us}. 
Bronnikov \textit{et al.} showed
that a wormhole that has the metric identical to that of the Ellis 
wormhole metric but with 
electrically charged dust with negative energy 
density~\cite{Shatskiy:2008us,Novikov:2012uj,Konoplya:2016hmd}
is linearly stable against spherical and axial perturbations~\cite{Bronnikov:2013coa}.
Their result shows clearly that instability of wormholes depends not only on
the metric but also on the properties of the matter field
that is a source of the metric.

Light rays passing through the Ellis wormhole have been studied by Ellis~\cite{Ellis_1973} 
and the gravitational lensing effects investigated by Perlick~\cite{Perlick_2004_Phys_Rev_D}.
Gravitational lensing by the Ellis wormhole in
the strong-field regime 
has been studied~\cite{Chetouani_Clement_1984,Muller:2008zza,Perlick_2004_Phys_Rev_D,Nandi_Zhang_Zakharov_2006,Dey_Sen_2008}.
The visualization of the Ellis wormhole~\cite{Muller_2004},  
time delay of light rays~\cite{Nakajima:2014nba},
particle collision at a throat~\cite{Tsukamoto:2014swa}, 
images surrounded by optically thin dust~\cite{Ohgami:2015nra},
and effect of a plasma on a shadow~\cite{Perlick:2015vta} have also been investigated.

Recently, the upper bound of the number density of the Ellis wormhole was estimated.
Takahashi and Asada~\cite{Takahashi_Asada_2013} have presented the upper bound of the number density $\leq 10^{-4}h^{3}$Mpc$^{-3}$
for a throat radius parameter $a$ to be in the astronomical scale,
i.e., $10^{1}\leq a\leq10^{4}$pc
by using the Sloan Digital Sky Survey Quasar Lens Search~\cite{Inada_Oguri_Shin_et_al_2012}, 
which has the largest quasar lens sample in the Sloan Digital Sky Survey~\cite{York_Adelman_Anderson_et_al_2000}. 
Yoo \textit{et al.}~\cite{Yoo_Harada_Tsukamoto_2013} have given 
the upper bound of the number density $\leq 10^{-9}$AU$^{-3}$ for the
daily-life scale throat $a \simeq 1$cm 
with the femto-lensing effect of the gamma-ray bursts~\cite{Barnacka_Glicenstein_Moderski_2012} 
by the data of the Fermi Gamma-Ray Burst Monitor~\cite{Meegan_Lichti_Bhat_et_al_2009}.

The Ellis wormhole 
has a vanishing Arnowitt-Deser-Misner (ADM)
mass with a gravitational potential that is asymptotically proportional to $1/r^{2}$, 
where $r$ is an areal radial coordinate.  
Abe pointed out that the light curves in the Ellis wormhole spacetime 
have characteristic gutters 
both before and after a peak~\cite{Abe_2010}. 
We can distinguish the light curves of the Ellis wormhole 
from those of usual massive 
objects such as planets, stars, black holes, and galaxies 
with the gravitational lensing effects under the weak-field approximation~\cite{Chetouani_Clement_1984,Muller:2008zza,
Abe_2010,Toki_Kitamura_Asada_Abe_2011,Tsukamoto_Harada_2013,Yoo_Harada_Tsukamoto_2013,Takahashi_Asada_2013,Perlick_2004_Phys_Rev_D,Lukmanova_2016}.

Since the behavior of the gravitational lensing in the weak
gravitational field is solely determined by the line element 
in the asymptotic region of the spacetime,
we cannot find the difference 
between the wormholes with a positive mass and the usual massive objects 
under the weak-field approximation.
We also cannot distinguish the Ellis wormhole 
from other exotic objects with an effective potential 
asymptotically proportional to $1/r^{2}$ by their gravitational lensing 
effects under the weak-field approximation~\cite{Nakajima:2014nba,Kitamura_Nakajima_Asada_2013,Izumi_Hagiwara_Nakajima_Kitamura_Asada_2013,Tsukamoto_Harada_2013}.

How do we distinguish a traversable wormhole with a positive mass from other massive objects? 
The wormhole has a throat on a region in a strong gravitational field 
and light rays can pass through the throat from the other side 
because of the nonexistence of the event horizon.
It may be good to pay attention to phenomena near a wormhole throat for answering the question.

In this paper, we give a method to find wormholes that have a
positive or zero mass with their gravitational lensing effects.
For simplicity, we concentrate on the Ellis wormhole 
with a vanishing ADM mass.
We investigate the light curves due to light rays coming 
from another asymptotic region through the throat.

This paper is organized as follows.
In Sec.~II we review gravitational lenses of light rays that pass through a wormhole throat 
and then we present the light curves of light rays emitted by a moving source. 
The latter is our original result.
In Sec.~III we consider light curves of light rays that do not pass through a wormhole throat
to compare the shapes of light curves.
We consider microlensing in a usual gravitational lens configuration under the weak-field approximation and retrolensing 
in the Ellis wormhole, an ultrastatic Schwarzschild-like wormhole, and the Schwarzschild spacetime, where the term ``ultrastatic'' means
that there is a timelike Killing vector that is hypersurface orthogonal
and of constant norm.
In Sec.~IV we discuss our results and conclude the paper.
In this paper we use the units in which the light speed and Newton's constant are unity.

\section{Gravitational lensing of a light that passes through a wormhole throat} 
In this section, we review gravitational lensing with an exact lens equation~\cite{Perlick_2004_Phys_Rev_D} in the Ellis wormhole spacetime
and investigate the light curves of a light ray that passes through a throat.
The line element in the Ellis wormhole spacetime is given by
\begin{eqnarray}
ds^{2}
=-dt^{2}+dr^{2}+(r^{2}+a^{2})(d\theta^{2}+\sin^{2}\theta d\phi^{2}), 
\end{eqnarray}
where $a$ is a positive constant and the coordinates are defined in the range $-\infty<t<\infty$, $-\infty<r<\infty$, 
$0\le \theta \le \pi$,
and $0 \leq \phi <2\pi$.\footnote{
The radial coordinate $r$ is a proper radial distance from a wormhole throat 
denoted as $l$ in Ref.~\cite{Morris_Thorne_1988}.
}
Note that the Ellis wormhole is not only static 
but also ultrastatic, i.e., $g_{tt}=\mathrm{const}$.
To consider geodesics, we
assume $\theta=\pi/2$ without loss of 
generality because of spherical symmetry.

\subsection{Initial conditions}
We consider that a past-oriented null geodesic starts from an observer.
We impose the initial conditions of the null geodesic 
\begin{eqnarray}\label{eq:t}
&&\left. t \right|_{\lambda=0}=0,\\
&&\left. r\right|_{\lambda=0}=r_{O}<0,\\ 
&&\left. \phi \right|_{\lambda=0}=0,\\\label{eq:dt_ds}
&&\left. \frac{dt}{d\lambda}\right|_{\lambda=0}=-1,\\
&&\left. \frac{dr}{d\lambda}\right|_{\lambda=0}=\cos \Theta,\\\label{eq:dphi_ds}
&&\left. \frac{d\phi}{d\lambda} \right|_{\lambda=0}=\frac{\sin \Theta}{\sqrt{r_{O}^{2}+a^{2}}}, 
\end{eqnarray}
where $\lambda$ is an affine parameter 
and we have set $\lambda=0$ at $r=r_{O}$ and $\phi=0$, the
location of the observer,
and $\Theta$ is the colatitude coordinate on the observer's sky.

Since the Ellis wormhole spacetime is 
static and spherically symmetric,
we have conserved energy
\begin{eqnarray}\label{eq:conserved_energy}
\frac{dt}{d\lambda}=-1
\end{eqnarray}
and conserved angular momentum 
\begin{eqnarray}\label{eq:conserved_angular_momentum}
(r^{2}+a^{2})\frac{d\phi}{d\lambda}=\sqrt{r_{O}^{2}+a^{2}}\sin \Theta
\end{eqnarray}
along the null geodesic from
Eqs.~(\ref{eq:dt_ds}) and (\ref{eq:dphi_ds}).

From Eqs.~(\ref{eq:conserved_energy}) and (\ref{eq:conserved_angular_momentum}) and $k^{\mu}k_{\mu}=0$, where $k^{\mu}$ is the photon wave number, we obtain
\begin{eqnarray}\label{eq:geodesic_equation}
(r^{2}+a^{2}) \left( \frac{dr}{d\lambda} \right)^{2}
=r^{2}+a^{2}-(r_{O}^{2}+a^{2}) \sin^{2}\Theta.
\end{eqnarray}
We can check easily that Eq.~(\ref{eq:geodesic_equation}) satisfies the initial condition~(\ref{eq:t})-(\ref{eq:dphi_ds}) at $\lambda=0$.

\subsection{Configuration of the gravitational lens}
\begin{figure}[htbp]
\begin{center}
\includegraphics[width=60mm]{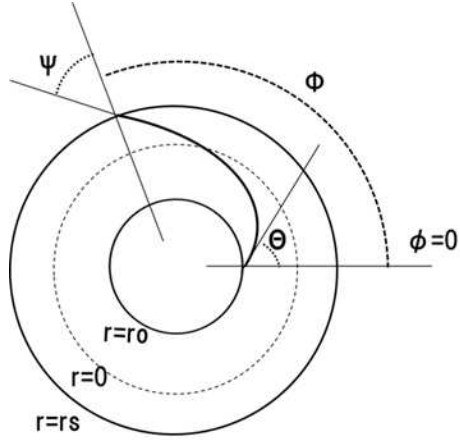}
\end{center}
\caption{
The configuration of the gravitational lens. 
A past-oriented null geodesic starts from an observer at $r=r_{O}<0$ and
 $\phi=0$ in the direction 
with the colatitude coordinate $\Theta$ in the observer's sky.
It passes through a throat at $r=0$ 
and it reaches into a source at $r=r_{S}>0$ and $\phi=(\Phi \bmod 2\pi)$, 
where $\Phi$ is the azimuthal angle swept out by the null geodesic from the observer to the source. 
$\Psi$ is an angle between the $r$ axis and the tangent of the past-oriented null geodesic at the source.
}
\label{Fig1}
\end{figure}

We concentrate on a past-oriented null geodesic that passes through a throat at $r=0$ 
and it reaches a source at $\lambda=\lambda_{S}>0$, where
$t=-\lambda_{S}<0$, $r=r_{S}>0$, and $\phi=(\Phi \bmod 2\pi)$; $\Phi$ is
the azimuthal angle swept out by
the null geodesic from the observer to the source. 
The configuration of the gravitational lens is depicted in Fig~\ref{Fig1}.
We define $\Psi$ as an angle between the $r$ axis and the tangent of the
past-oriented null geodesic at the source, where $\lambda=\lambda_{S}$.
From Eq.~(\ref{eq:conserved_energy}), $k^{\mu}k_{\mu}=0$, and the definition of $\Psi$, we obtain 
\begin{eqnarray}
&&\left. \frac{dt}{d\lambda}\right|_{\lambda=\lambda_{S}}=-1,\\
&&\left. \frac{dr}{d\lambda}\right|_{\lambda=\lambda_{S}}=\cos \Psi,\\ 
&&\left. \frac{d\phi}{d\lambda} \right|_{\lambda=\lambda_{S}}=\frac{\sin \Psi}{\sqrt{r_{S}^{2}+a^{2}}}.\label{eq:Psi}
\end{eqnarray}
Since the angular momentum $(r^{2}+a^{2})d\phi/d\lambda$ is constant along the geodesic, 
from Eqs.~(\ref{eq:conserved_angular_momentum}) and (\ref{eq:Psi}), we obtain 
\begin{eqnarray}\label{eq:Psi_Theta_relation}
\sqrt{r_{S}^{2}+a^{2}}\sin\Psi
=\sqrt{r_{O}^{2}+a^{2}}\sin\Theta.
\end{eqnarray} 

Since the right-hand side of Eq.~(\ref{eq:geodesic_equation}) should be
non-negative in the region $r_{O}\leq r\leq r_{S}$, at least, 
\begin{eqnarray}
-\delta_{I} \leq \Theta \leq \delta_{I}
\end{eqnarray}
must be satisfied, where
\begin{eqnarray}
\delta_{I}\equiv \arcsin \sqrt{\frac{a^{2}}{r_{O}^{2}+a^{2}}}.
\end{eqnarray}

\subsection{Angular diameter distance and luminosity distance}
We define angular diameter distance and luminosity distance in a static spherically symmetric spacetime with the line element given by
\begin{equation}
ds^{2}=g_{tt}(r)dt^{2}+g_{rr}(r)dr^{2}+g_{\theta\theta}(r)(d\theta^{2}+\sin^{2}\theta d\phi^{2}). 
\end{equation}
The Ellis wormhole spacetime has $g_{tt}(r)=-1$, $g_{rr}(r)=1$, and $g_{\theta\theta}(r)=r^{2}+a^{2}$.
We follow the definitions by Perlick~\cite{Perlick_2004_Phys_Rev_D}. 

\begin{figure}[htbp]
\begin{center}
\includegraphics[width=80mm]{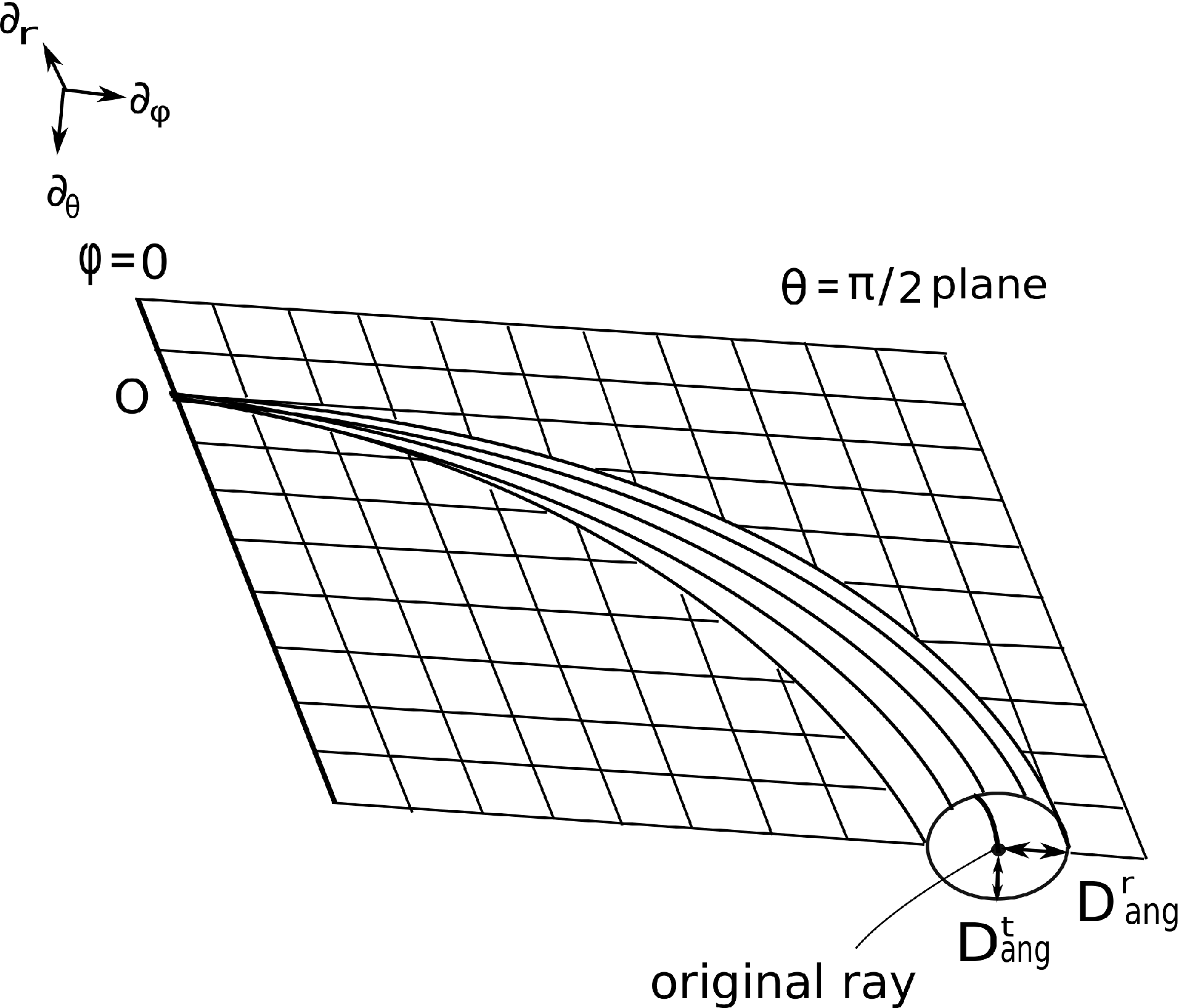}
\end{center}
\caption{
The radial angular diameter distance $D^{r}_{ang}$ and the tangential angular diameter distance $D^{t}_{ang}$.
}
\label{Figbundle}
\end{figure}

A light ray is given by a solution $(\Theta, \Phi)$ of a lens equation,
\begin{eqnarray}
\mathcal{F}(\Theta, \Phi)=0.
\end{eqnarray}
We obtain an infinitesimally neighboring light ray as a solution $(\Theta+d\Theta, \Phi+d\Phi)$
of the lens equation which can be expressed as
\begin{eqnarray}
\frac{\partial \mathcal{F}}{\partial \Theta}(\Theta, \Phi)+\frac{\partial \mathcal{F}}{\partial \Phi}(\Theta, \Phi)\frac{d\Phi}{d\Theta}=0.
\end{eqnarray}
The radial angular diameter distance $D^{r}_{ang}$ is defined as 
\begin{equation}\label{eq:Drang_general}\label{eq:Drang_def}
D^{r}_{ang}=\sqrt{g_{\theta\theta}(r_{S})} \cos\Psi \frac{d\Phi}{d\Theta},
\end{equation}
where $\sqrt{g_{\theta\theta}(r_{S})} \cos\Psi d\Phi$ is the distance between the original light ray and the infinitesimally neighboring light ray
in the direction perpendicular to the original light ray
shown in Fig.~\ref{Figbundle}.

We consider an infinitesimally neighboring ray 
made by an infinitesimal rotation with an infinitesimal angle $d\theta$ around the axis $\phi=0$ on the equatorial plane $\theta=\pi/2$ 
that is generated by the Killing vector $\textrm{\boldmath $K$}=\sin\phi \partial_{\theta}$.
The distance between the original ray at $r=r_{S}$, $\theta=\pi/2$, and $\phi=\Phi$ and the infinitesimally neighboring ray
is given by $\sqrt{g_{\theta\theta}(r_{S})}\sin \Phi d\theta$.
The tangential angular diameter distance is defined as the distance over
the observer's angle $\sin \Theta d\theta$
between the original ray and the infinitesimally neighboring ray,
\begin{equation}\label{eq:Dtang_def}
D^{t}_{ang}=\sqrt{g_{\theta\theta}(r_{S})}\frac{\sin \Phi}{\sin \Theta}.
\end{equation}
The angular diameter distance $D_{ang}$ is given by 
\begin{eqnarray}\label{eq:Dang}
D_{ang}=\sqrt{\left| D^{r}_{ang} D^{t}_{ang} \right|}.
\end{eqnarray}

From a well-known reciprocal relation between the luminosity distance $D_{lum}$ and the angular diameter distance $D_{ang}$
\begin{equation}
D_{lum}=(1+z_{red})^{2}D_{ang},
\end{equation}
where $z_{red}=\sqrt{g_{tt}(r_{O})/g_{tt}(r_{S})}-1$ is the redshift,
we obtain the luminosity distance $D_{lum}$ as
\begin{equation}\label{eq:Dlum_ang}
D_{lum}=D_{ang}
\end{equation}
in the Ellis wormhole spacetime because of $z_{red}=0$.

\subsection{Primary and secondary images}
It is well known that 
an infinite number of images appear near a light sphere because of the strong gravitational field~\cite{Darwin_1959,Perlick_2004_Phys_Rev_D}.
In this paper, 
we define a primary image as 
the image due to the light ray that sweeps the smallest winding angle $|\Phi|$.
We similarly define secondary, tertiary, and
quaternary images according to the winding 
angle $|\Phi|$ swept out by the light ray.
In the following, we mainly discuss the primary and secondary images.
Only when we cannot ignore the effects of higher-order images will we discuss them.
We concentrate on the images of the source in the range $0 \le \phi < \pi$ 
because we can obtain images of the source in the range $\pi \le \phi <
2\pi$ from symmetry with respect to $\phi=\pi$.

\begin{figure}[htbp]
\begin{center}
\includegraphics[width=80mm]{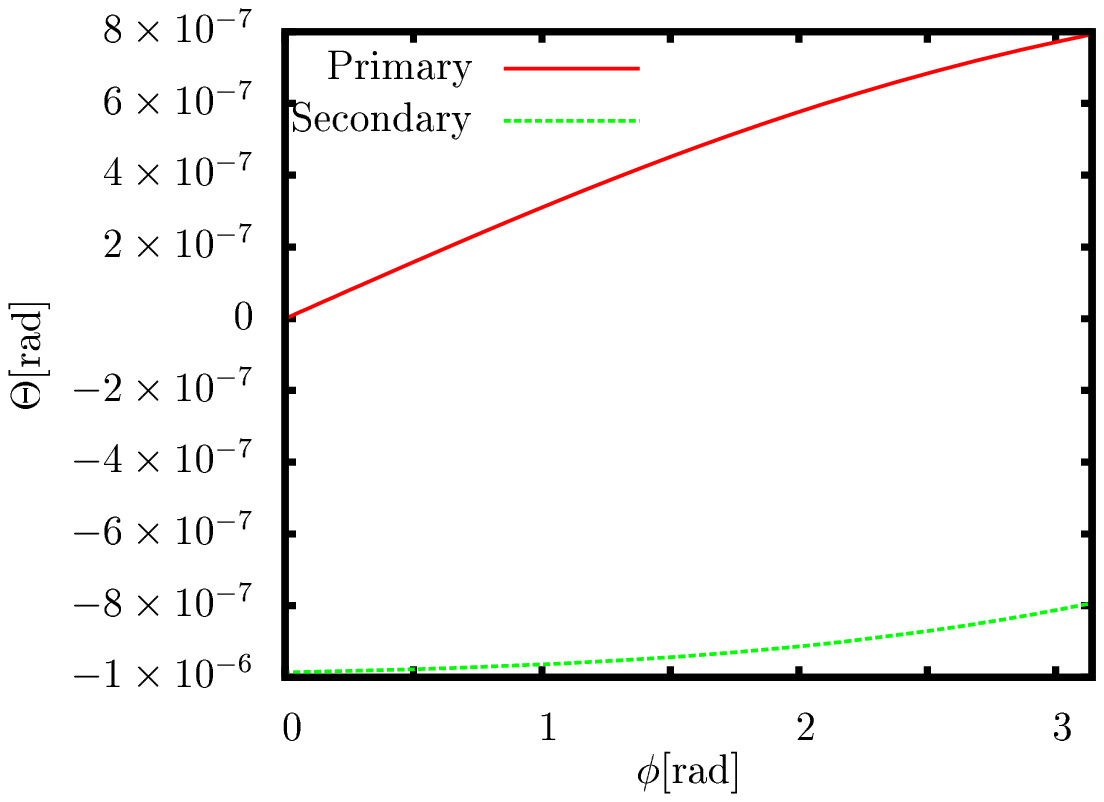}
\end{center}
\caption{$\Theta$ against the azimuthal angle $\phi$. We set $a=10^{-2}$ pc and $r_{S}=-r_{O}=10$ kpc.
Note that $\delta_{I}\sim 10^{-6}$ rad in this case.
Solid (red) and broken (green) curves denote the primary and secondary images, respectively.
}
\label{Fig2}
\end{figure}

From Eqs.~(\ref{eq:conserved_angular_momentum}) and (\ref{eq:geodesic_equation}), $\Phi$ is obtained as
\begin{eqnarray}\label{eq:Phi}
\Phi
&=&\int^{r_{S}}_{r_{O}}\frac{\sqrt{r_{O}^{2}+a^{2}}\sin\Theta dr}
{\sqrt{r^{2}+a^{2}}\sqrt{r^{2}+a^{2}\cos^{2}\Theta-r_{O}^{2}\sin^{2}\Theta}} \nonumber\\
&=&hI,
\end{eqnarray}
where $0 \leq  h\equiv \sin \Theta/\sin \delta_{I} \leq 1$ and $I$ is defined by
\begin{eqnarray}
I
&\equiv& F \left( \arctan \frac{-r_{S}}{r_{O}\sqrt{1-h^{2}}\tan\delta_{I}}, h \right) \nonumber\\
&+& F \left( \arctan \frac{1}{\sqrt{1-h^{2}}\tan\delta_{I}}, h \right)\
\end{eqnarray}
and $F(\varphi, k)$ is the elliptic integral of the first kind defined as 
\begin{eqnarray}
F(\varphi, k)
\equiv 
\int^{\varphi}_{0} \frac{d\theta}{\sqrt{1-k^{2}\sin^{2}\theta}}.
\end{eqnarray}
Equation~(\ref{eq:Phi}) is a lens equation that we solve.
$\Phi$ is monotonically increasing with respect to $\Theta$ and changes from $0$ to $\infty$ 
as $\Theta$ increases from $0$ to $\delta_{I}$.
$\Theta$ as a function of the azimuthal angle $\phi$ is plotted in Fig.~\ref{Fig2}.

\begin{figure}[htbp]
\begin{center}
\includegraphics[width=80mm]{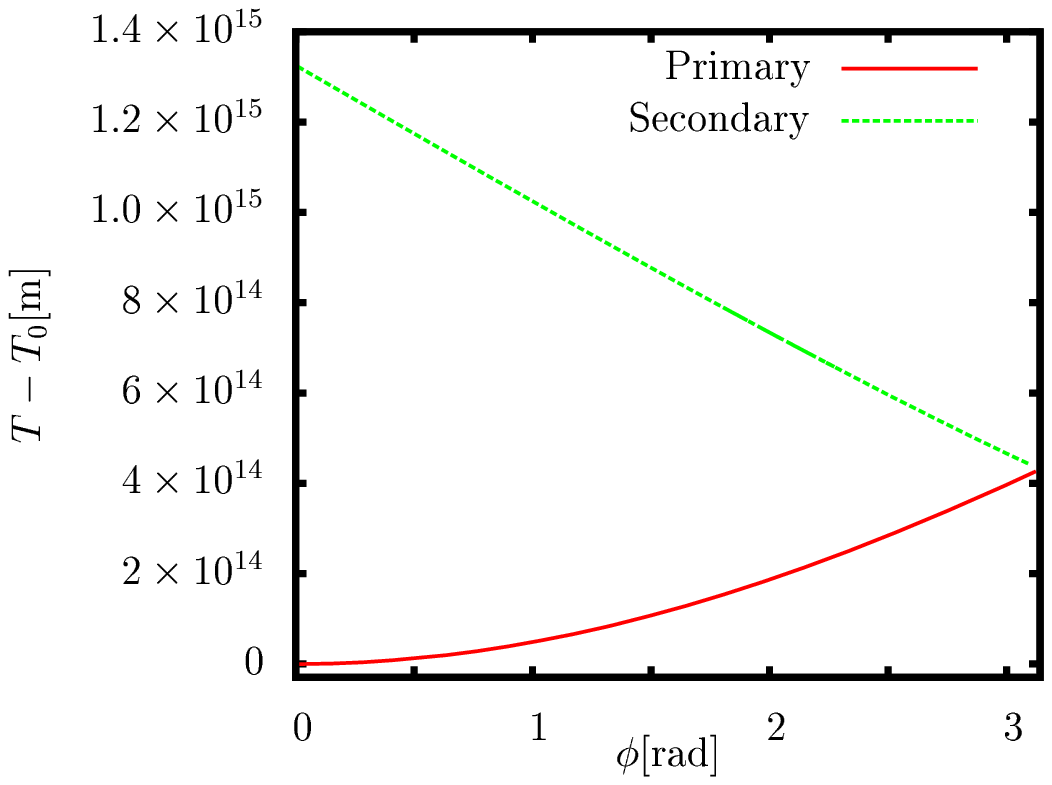}
\end{center}
\caption{The relationship between the travel time excess $T-T_{0}$ and the azimuthal angle $\phi$. We put $a=10^{-2}$ pc and $r_{S}=-r_{O}=10$ kpc.
Solid (red) and broken (green) curves denote the primary and secondary images, respectively.
}
\label{Fig3}
\end{figure}

From Eq.~(\ref{eq:geodesic_equation}) and $\lambda_{S}=T$, 
the travel time $T$ is given by 
\begin{eqnarray}
T
&=&\int^{r_{S}}_{r_{O}}
\frac{\sqrt{r^{2}+a^{2}} dr}
{\sqrt{r^{2}+a^{2}\cos^{2}\Theta-r_{O}^{2}\sin^{2}\Theta}} \nonumber\\
&=&aI +J,
\end{eqnarray}
where 
$J$ is a regular integral given by
\begin{eqnarray}
J
\equiv \int^{r_{S}}_{r_{O}}\frac{r^{2}dr}{\sqrt{r^{2}+a^{2}}\sqrt{r^{2}+a^{2} \left(1-h^{2}\right)}}.
\end{eqnarray}
$J$ and $T$ monotonically increase with respect to $\Theta$ and
change from 
$r_{S}-r_{O}-a\arcsin (r_{S}/a)+a\arcsin (r_{O}/a)$ to $\sqrt{r_{S}^{2}+a^{2}}+\sqrt{r_{O}^{2}+a^{2}}-2a$ 
and from $T_{0}\equiv r_{S}-r_{O}$ to $\infty$, respectively, as $\Theta$ increases from $0$ to $\delta_{I}$.
Figure~\ref{Fig3} shows $T-T_{0}$ as a function of $\phi$. 
The travel time of the secondary image is always longer than the primary image.
We note that $T-T_{0}\sim a \left| \Phi \right|+$  const for $\left| \Phi \right| \gtrsim \pi$, 
i.e., if the light ray winds around the wormhole.

From Eqs.~(\ref{eq:Psi_Theta_relation}), (\ref{eq:Drang_def}), and (\ref{eq:Phi}), the radial angular diameter distance is given by 
\begin{eqnarray}
D^{r}_{ang}
=&&\sqrt{r_{S}^{2}+a^{2}\cos^{2}\Theta-r_{O}^{2}\sin^{2}\Theta }\sqrt{r_{O}^{2}+a^{2}} \cos \Theta \nonumber\\
&&\times \int^{r_{S}}_{r_{O}}\frac{\sqrt{r^{2}+a^{2}}dr}{\left( \sqrt{r^{2}+a^{2}\cos^{2}\Theta-r_{O}^{2}\sin^{2}\Theta}\right)^{3}} \nonumber\\
=&&\frac{\sqrt{r_{S}^{2}+a^{2}(1-h^{2})}\cos\Theta}{(1-h^{2})\sin\delta_{I}} \nonumber\\
&&\times \left[ 2E\left( h \right) -E\left( \arcsin \frac{a}{\sqrt{a^{2}+r_{S}^{2}}}, h \right) -E\left( \delta_{I}, h \right)  \right. \nonumber\\
&& \left. +\frac{\sin^{2}\Theta}{\tan\delta_{I}\cos\Theta} 
+\frac{ar_{S}h^{2}}{(a^{2}+r_{S}^{2})\sqrt{1-\frac{a^{2}h^{2}}{a^{2}+r_{S}^{2}}}} \right], 
\end{eqnarray}
where $E(\varphi,k)$ is the elliptic integral of the second kind that is defined as
\begin{eqnarray}
E(\varphi,k)\equiv \int^{\varphi}_{0} \sqrt{1-k^{2}\sin^{2}\theta}d\theta
\end{eqnarray}
and $E(k)$ is the complete elliptic integral of the second kind defined as $E(k)\equiv E(\pi/2,k)$.
$D^{r}_{ang}$ monotonically increases with respect to $\Theta$ and changes from 
$\sqrt{r_{S}^{2}+a^{2}}(\pi/2-\delta_{I}+\arctan (r_{S}/a))/\sin \delta_{I}$ to $\infty$ as $\Theta$ increases from $0$ to $\delta_{I}$.

From Eqs.~(\ref{eq:Dtang_def}) and (\ref{eq:Phi}), we obtain the tangential angular diameter distance as 
\begin{eqnarray}
D^{t}_{ang}
=\frac{\sqrt{r_{S}^{2}+a^{2}}}{\sin\Theta}\sin hI.
\end{eqnarray}
The parity of an image is determined
by the sign of $D^{r}_{ang}D^{t}_{ang}$.
A primary image has even parity $D^{r}_{ang}D^{t}_{ang}>0$ and
a secondary image has odd parity $D^{r}_{ang}D^{t}_{ang}<0$.  
Thus, the primary image is a normal image and the secondary image is a mirror-symmetric image of the source.
A primary image with $\Theta=\Phi=0$ is 
not distorted at all in shape
since the radial and tangential angular diameter distances are the same:
\begin{eqnarray}
D^{r}_{ang}
=D^{t}_{ang}
=\frac{\sqrt{r_{S}^{2}+a^{2}}}{\sin \delta_{I}} \left( \frac{\pi}{2}-\delta_{I}+\arctan \frac{r_{S}}{a} \right). \nonumber\\
\end{eqnarray}

\begin{figure}[htbp]
\begin{center}
\includegraphics[width=80mm]{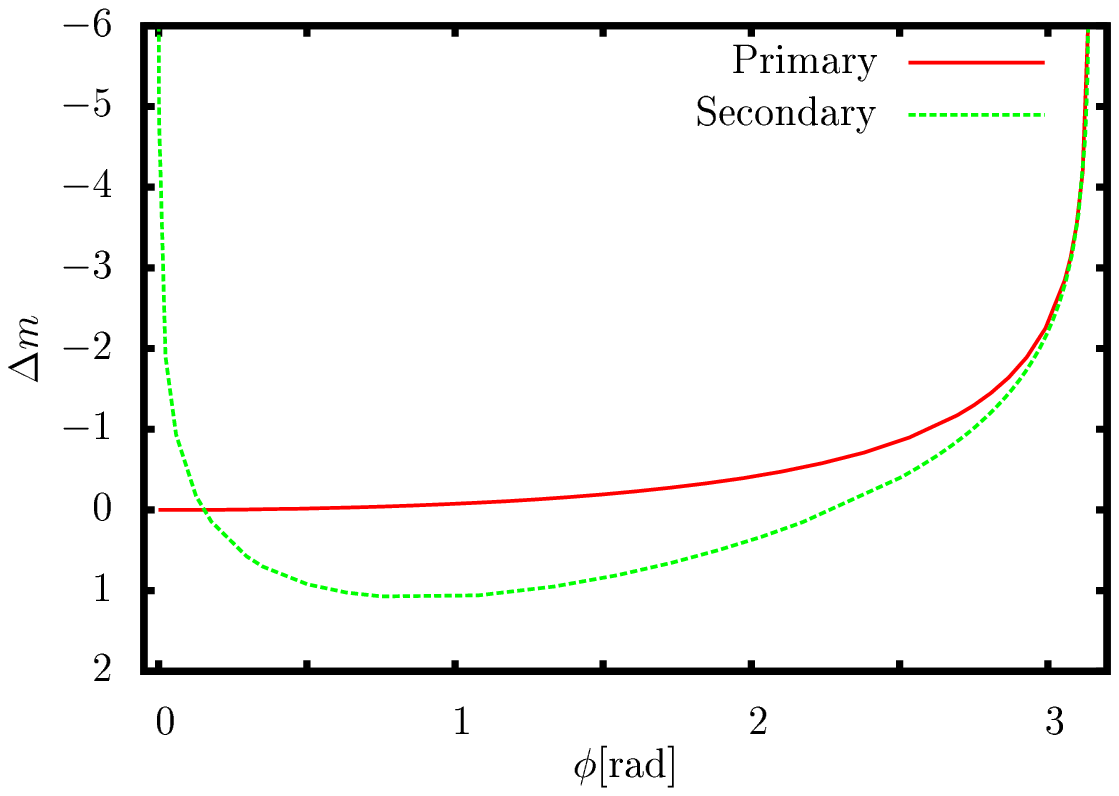}
\end{center}
\caption{The relationship between the 
relative magnitude $\Delta m$ of an observed image
and the azimuthal angle $\phi$ of the source. The relative magnitude is 
defined by the magnitude relative to the fiducial 
image with $\Theta=\Phi=0$. 
We set $a=10^{-2}$ pc and $r_{S}=-r_{O}=10$ kpc.
Solid (red) and broken (green) curves denote the primary and secondary images, respectively.
}
\label{Fig4}
\end{figure}

We define the relative magnitude $\Delta m$ of an observed image with respect to the image with $\Theta=\Phi=0$. 
That is, we
choose the image with $\Theta=\Phi=0$ as a fiducial image
for which the light ray of the image does not bend at all
so that the observer does not see any gravitational lensing effects.
By using Eqs. (\ref{eq:Dang}) and (\ref{eq:Dlum_ang}), 
the relative magnitude of an observed image is defined as
\begin{eqnarray}
\Delta m
\equiv
2.5\log_{10} \left| D^{r}_{ang}D^{t}_{ang} \right| +m_{0},
\end{eqnarray}
where $m_{0}$ is a constant given by 
\begin{eqnarray}
m_{0}\equiv -5\log_{10} \left[ \frac{\sqrt{r_{S}^{2}+a^{2}}}{\sin \delta_{I}} \left( \frac{\pi}{2}-\delta_{I}+\arctan \frac{r_{S}}{a} \right) \right]. \nonumber\\
\end{eqnarray}
The relative magnitude $\Delta m$ is plotted as a function of the
azimuthal angle $\phi$ in Fig.~\ref{Fig4}. 
$\Delta m$ of the primary image diverges at $\phi=\pi$, 
while that of the secondary image diverges at both $\phi=0$ and $\phi=\pi$.
The divergence occurs 
since a thin bundle of light rays collapses on 
the axis of symmetry, where $D^{t}_{ang}=0$.

\subsection{Light curves}
\begin{figure}[htbp]
\begin{center}
\includegraphics[width=80mm]{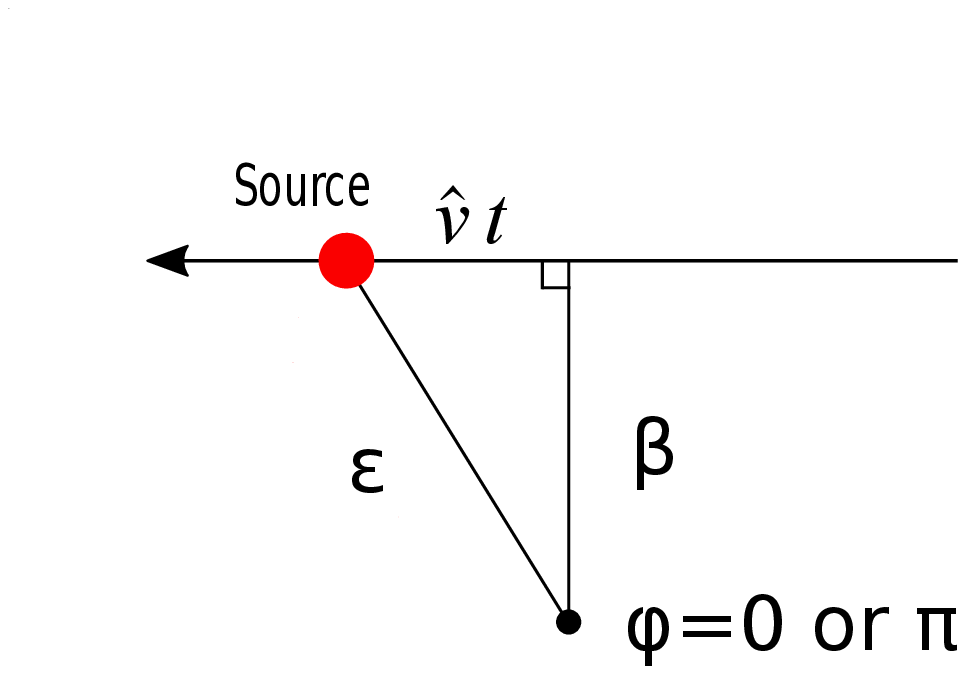}
\end{center}
\caption{
The motion of a source with respect to the axis $\phi=0$ or $\pi$, 
projected on the source plane.
The source moves with a velocity $\hat{v}$
either near the half-line axis $\phi=0$ or $\pi$ on the source plane.
The closest separation between the source and $\phi=0$ or $\pi$ is denoted by $\beta$.}
\label{Fig5}
\end{figure}

In this subsection, we consider light curves due to light rays coming from the other side of the throat.
We assume $a=10^{-2}$ pc and $r_{S}=-r_{O}=10$ kpc for reference. 
In this case, $\delta_{I}\sim 10^{-6}$ rad.
We consider two cases where a source moves with a velocity $\hat{v}=3\times 10^{-15}$ rad/s 
on the source plane near 
half-line axes $\phi=0$ and $\phi=\pi$. 
We denote the closest separation between the source and 
the axis $\phi=0$ by $\beta$.
We also denote the closest separation between the source and 
the axis $\phi=\pi$ by the same symbol $\beta$.
Figure~\ref{Fig5} illustrates the 
situation projected on the source plane.

If the relative velocity of the source with respect to the lens and 
the observer is sufficiently small, 
we can apply the static gravitational lens system discussed 
in Sec.~II for a lensing event. In other words, we consider
the change of the source position
$\phi=\phi\left|_{\lambda=\lambda_{s}}\right.$
but not other effects such as the kinematic Doppler effect.
We do not consider the effect of parallax on the light curves either.
Note that even if 
the source is not in the equatorial plane $\theta=\pi/2$,
we can calculate the luminosity by 
redefining the equatorial plane so that the source is on it. 

\subsubsection{Source passing by $\phi=\pi$}
\begin{figure}[htbp]
\begin{center}
\includegraphics[width=80mm]{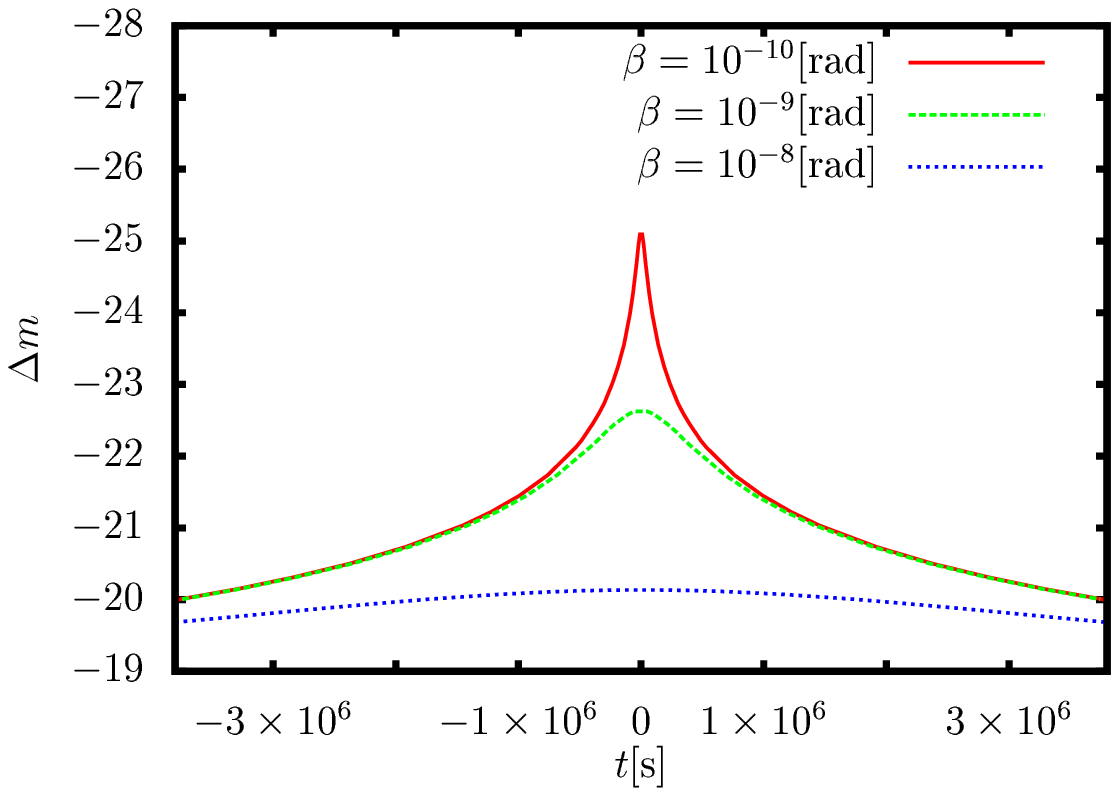}
\end{center}
\caption{
The light curves of the primary image with $\Phi=\pi-\epsilon$. 
The source is located at $\phi=\pi-\epsilon$, where 
$\epsilon$ is given by $\epsilon= \sqrt{\beta^{2}+\hat{v}^{2}t^{2}}$ with 
$\hat{v}=3\times 10^{-15}$ rad/s.
Solid (red), broken (green), and dotted (blue) curves denote the light curves 
with $\beta=10^{-10}$, $10^{-9}$, and $10^{-8}$ rad, respectively.}
\label{Fig6}
\end{figure}

Here, we assume that the source passes
by $\phi=\pi$, i.e., $\phi=\pi-\epsilon$, where 
$\epsilon (\ll 1)$ is given by
$\epsilon= \sqrt{\beta^{2}+\hat{v}^{2}t^{2}}$.
Note that we have set the time $t=0$ when the source is at the closest separation
$\beta$ to the axis $\phi=\pi$.
Thus, the coordinate $\phi$ of the source is expressed by
\begin{eqnarray}
\phi = \pi-\sqrt{\beta^{2}+\hat{v}^{2}t^{2}}.
\end{eqnarray}
The light curves of the primary image with $\Phi=\pi-\epsilon$ are plotted in Fig.~\ref{Fig6}.
The peak magnitude of the light curve depends on $\beta$.

As inferred from Fig.~\ref{Fig4},
the light curve of the secondary image with $\Phi=-\pi-\epsilon$ 
is very similar to the primary image with 
$\Phi=\pi-\epsilon$ both in shape and magnitude 
because of symmetry with respect to $\phi=\pi$.
There is of course the difference of the travel time between the two light curves and hence 
their peaks, which is given by $2a\beta$.
For example, if $\beta$ is $10^{-8}$ rad, 
the time difference is given by $2\times 10^{-2}$s.
Since the time interval of imaging in the current 
microlensing observation is much longer than this, 
these two light curves would not be separated but observed as
a superposed light curve with a single peak. In such a case, 
the observed light curve is twice as bright as that of the primary image. On the other hand, 
with a different set of parameter values, it would be possible to 
separate the two light curves and observe a double-peaked light curve.

Tertiary and higher-order images are much fainter than the primary and
secondary images~\cite{Perlick_2004_Phys_Rev_D}. 
So, when the source is 
at $\phi=\pi$, a pair of the primary image with 
$\Phi=\pi$ and the secondary image with $\Phi=-\pi$ constitutes the brightest 
and innermost Einstein ring among an infinite number of Einstein rings.

\subsubsection{Source passing by $\phi = 0$}
Next we assume that the source passes by $\phi=0$, i.e., 
$\phi=\epsilon$, where $\epsilon(\ll 1)$ is given by 
$\epsilon = \sqrt{\beta^{2}+\hat{v}^{2}t^{2}}$.
Then, the azimuthal angle $\phi$ of the source is given by
\begin{eqnarray}
\phi = \sqrt{\beta^{2}+\hat{v}^{2}t^{2}}.
\end{eqnarray}

\begin{figure}[htbp]
\begin{center}
\includegraphics[width=80mm]{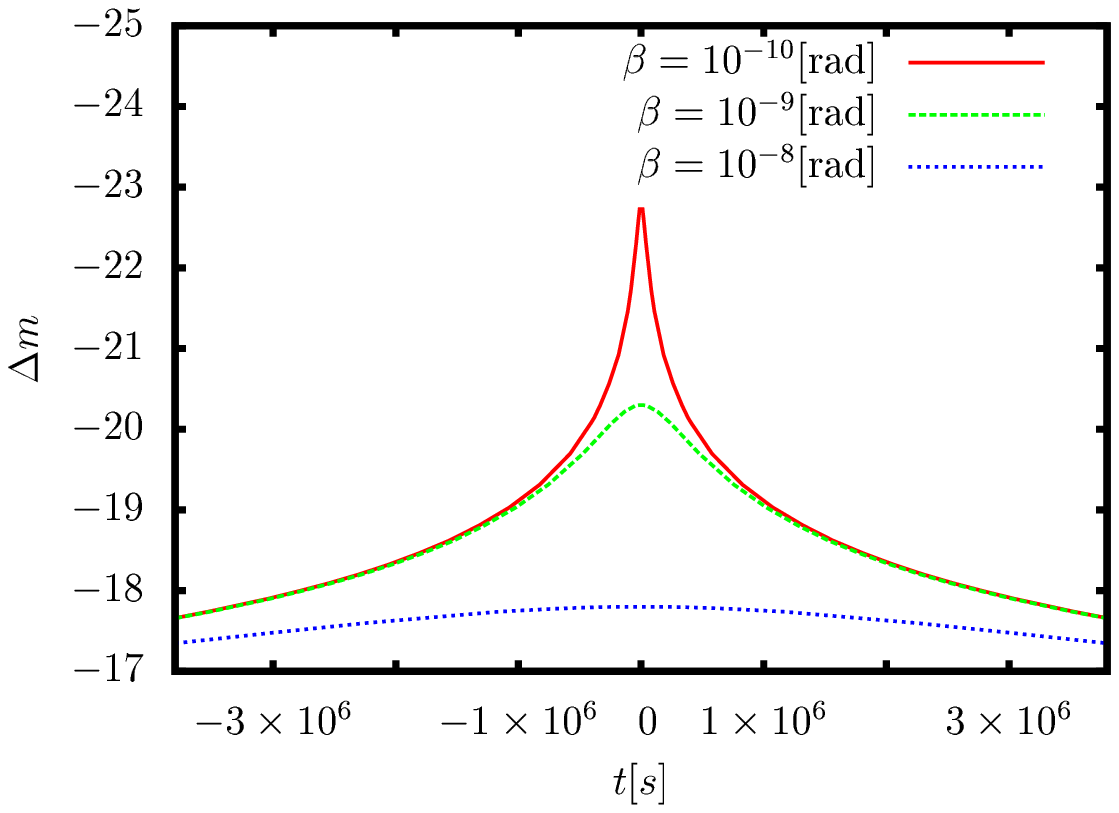}
\end{center}
\caption{
The light curves of the secondary image with $\Phi=-2\pi+\epsilon$.
The source is located at $\phi=\epsilon$, where 
$\epsilon$ is given by 
$\epsilon= \sqrt{\beta^{2}+\hat{v}^{2}t^{2}}$ with 
$\hat{v}=3\times 10^{-15}$ rad/s.
Solid (red), broken (green), and dotted (blue) curves denote the light curves 
with $\beta=10^{-10}$, $10^{-9}$, and $10^{-8}$ rad, respectively.}
\label{Fig7}
\end{figure}

If $\beta$ is sufficiently small, the 
relative magnitude $\Delta m$ of the primary image with $\Phi = \epsilon$ is almost $0$ and constant.
Figure~\ref{Fig7} shows the light curves of the secondary image with $\Phi = -2\pi+\epsilon$. 
The peak magnitude of the light curve of the secondary image depends on $\beta$.
Since the secondary image is much brighter than the primary image,
we can ignore the effect of the primary image on the light curves.

The light curves of the tertiary image with $\Phi=2\pi+\epsilon$ 
are very similar to that of the secondary image 
with $\Phi = -2\pi+\epsilon$ both in shape and magnitude
because of symmetry with respect to $\phi=0$.
These two images are a pair of relativistic images due to light rays 
that have passed around the light sphere of the wormhole 
with winding numbers $\pm 1$ along almost symmetric orbits.
It is known that 
the relativistic images of such a pair
have almost the same brightness~\cite{Darwin_1959}.
The time difference of the peaks of the light curves of the secondary and tertiary images is
given by $2a\beta$.
Thus, the observed light curve is twice as bright as 
that of the secondary image, if these two light
curves cannot be separated.

Quaternary and higher-order images are much 
dimmer than the secondary and tertiary images.
When the source is at $\phi = 0$,
a pair of the secondary image with $\Phi=-2\pi$ and the tertiary image with $\Phi = 2\pi$
constitutes the brightest and innermost Einstein ring among an infinite number of Einstein rings.

By comparing Figs.~\ref{Fig6} and~\ref{Fig7}, 
we notice that the observed light curve of the source passing by $\phi = \pi$
is much brighter than that by $\phi = 0$ if the 
closest separations to the axes are the same.

\section{Gravitational lensing of a light that does not pass through a wormhole throat} 
In this section, we investigate light curves in 
microlens and retrolens configurations and 
compare them with those obtained in Sec.~II.

\subsection{Deflection angle}
We investigate the deflection angle of a light ray that does not
pass through a wormhole throat in an ultrastatic Schwarzschild-like wormhole spacetime 
and review the ones in the Schwarzschild and Ellis spacetimes.

\subsubsection{An ultrastatic Schwarzschild-like wormhole}
We consider an ultrastatic Schwarzschild-like wormhole with a line element 
\begin{equation}
ds^{2}=-dt^{2}+\frac{d\rho^{2}}{1-\frac{2M_{w}}{\rho}} +\rho^{2}(d\theta^{2}+\sin^{2}\theta d\phi^{2}),
\end{equation}
where $M_{w}$ is the mass and $-\infty<t<\infty$, $2M_{w} \leq \rho<\infty$, $0 \leq \theta \leq \pi$, and $0 \leq \phi <2\pi$. 
This wormhole spacetime has three nonzero components of the Ricci tensor given by
\begin{eqnarray}
R_{\rho \rho}&=&-\frac{2M_{w}}{\left(1-\frac{2M_{w}}{\rho}\right)\rho^{3}}, \\
R_{\theta \theta}&=&\frac{R_{\phi\phi}}{\sin^{2}\theta}=\frac{M_{w}}{\rho}
\end{eqnarray}
but the Ricci scalar vanishes.
The throat is at $\rho=2M_{w}$. 
We introduce a coordinate $r$ taking a range $-\infty< r < \infty$ defined by
\begin{equation}
\frac{dr}{d\rho}=\pm \left( 1-\frac{2M_{w}}{\rho} \right)^{-\frac{1}{2}},
\end{equation}
where the upper (lower) sign is chosen when $r>0$ ($r<0$).
We can integrate it and obtain 
\begin{equation}
r=\pm \left[ \sqrt{\rho(\rho-2M_{w})}+2M_{w}\log \left( \sqrt{\frac{\rho}{2M_{w}}}+\sqrt{\frac{\rho}{2M_{w}}-1} \right) \right].
\end{equation}
Using $r$, the line element is rewritten as
\begin{equation}
ds^{2}=-dt^{2}+dr^{2} +\rho^{2}(r)(d\theta^{2}+\sin^{2}\theta d\phi^{2})
\end{equation}
and the throat is at $r=0$.
We can assume that $\theta=\pi/2$ without loss of generality.
The trajectory of a light ray is given by
\begin{equation}
\dot{r}^{2}-E^{2}+\frac{L^{2}}{\rho^{2}(r)}=0,
\end{equation}
where $\dot{\:}$ denotes a differentiation with respect to an affine parameter
and $E\equiv \dot{t}>0$ and $L\equiv \rho^{2}(r)\dot{\phi}$ 
are the conserved energy and angular momentum, respectively.
The wormhole has a light sphere at $r=0$ and it coincides with the throat.
We define the impact parameter of a light ray as $b\equiv L/E$. 
As long as we consider one light ray, we can assume that $L$ and $b$ are non-negative without loss of generality.
A light ray does not pass the throat if $b > 2M_{w}$ while it does if $b < 2M_{w}$. 

We consider the case of $b > 2M_{w}$.
A light ray comes from an infinity and it is deflected by a wormhole without passing the throat. 
The closest distance of a light ray from the wormhole is given by $\rho=b$.
The deflection angle $\alpha$ of a light ray is given by
\begin{eqnarray}\label{eq:da}
\alpha 
&=&2I_{1}-\pi \nonumber\\
&=&4\sqrt{\frac{b}{b+2M_{w}}}F\left( \arcsin \sqrt{\frac{b+2M_{w}}{2b}}, \sqrt{\frac{4M_{w}}{b+2M_{w}}} \right) -\pi, \nonumber\\
\end{eqnarray}
where $I_{1}$ is defined by
\begin{equation}
I_{1}\equiv \int^{\infty}_{b}\frac{bd\rho}{\sqrt{(\rho-2M_{w})\rho (\rho^{2}-b^{2})}}.
\end{equation}
Under the weak-field approximation $b \gg M_{w}$, the deflection angle becomes
\begin{equation}
\alpha 
=\frac{2M_{w}}{b}+ O\left( \left( \frac{M_{w}}{b} \right)^{2} \right).
\end{equation}
Notice that the leading term of the deflection angle is just half of the one in the Schwarzschild spacetime 
because the $(t,t)$ component of the metric tensor $g_{tt}=-1$ in the wormhole spacetime does not contribute to the deflection angle 
while $g_{tt}=-1+2M_{s}/\rho$, where $M_{s}$ is the ADM mass, in the Schwarzschild spacetime does.
The deflection angle diverges in a strong deflection limit $b \rightarrow b_{c}\equiv2M_{w}$.
We consider the deflection angle in the strong deflection limit $b
\rightarrow b_{c}$ in the following form~\cite{Bozza_2002,Tsukamoto:2016qro}:
\begin{equation}\label{eq:defstr}
\alpha(b)=-\bar{a}\log \left( \frac{b}{b_{c}}-1 \right) +\bar{b} +O((b-b_{c})\log (b-b_{c})),
\end{equation}
where $\bar{a}$ is a positive constant and $\bar{b}$ is a constant.
In ultrastatic spacetimes, a well-known method to obtain the deflection angle in the strong deflection limit investigated by Bozza \cite{Bozza_2002} does not work 
since several equations diverge. We will use an extended method for ultrastatic spacetimes to calculate it~\cite{Tsukamoto:2016qro}.
The integral $I_{1}$ can be rewritten as
\begin{equation}
I_{1}=\int^{1}_{0}f(z,b)dz,
\end{equation}
where $z$ is a variable defined as
\begin{equation}
z\equiv 1-\frac{b}{\rho}
\end{equation}
and $f(z,b)$ is given by 
\begin{equation}
f(z,b)\equiv \frac{\sqrt{b}}{\sqrt{c_{1}(b)z+c_{2}(b)z^{2}-2M_{w}z^{3}}},
\end{equation}
where $c_{1}(b)\equiv 2(b-2M_{w})$ and $c_{2}(b)\equiv -b+6M_{w}$.
Since in the strong deflection limit $b\rightarrow 2M_{w}$ we obtain 
$c_{1}\rightarrow 0$ and $c_{2}\rightarrow 4M_{w}$,
we notice that the leading order of the divergence of $f(z,b)$ is $z^{-1}$.
We separate the integral $I_{1}$ into 
\begin{equation}\label{eq:I}
I_{1}=I_{D}+I_{R}
\end{equation}
where $I_{D}$ is a divergent part and $I_{R}$ is a regular part .
We define the divergent part $I_{D}$ as 
\begin{eqnarray}
I_{D}(b)
&\equiv& \int^{1}_{0} f_{0}(z,b) dz \nonumber\\
&=&\frac{2\sqrt{b}}{\sqrt{-b+6M_{w}}}\log \frac{\sqrt{-b+6M_{w}}+\sqrt{b+2M_{w}}}{\sqrt{2(b-2M_{w})}},\nonumber\\
\end{eqnarray}
where 
\begin{equation}
f_{0}(z,b)\equiv \frac{\sqrt{b}}{\sqrt{c_{1}(b)z+c_{2}(b)z^{2}}}.
\end{equation}
In the strong deflection limit $b\rightarrow b_{c}=2M_{w}$, $I_{D}$ becomes
\begin{equation}\label{eq:ID}
I_{D}=-\frac{\sqrt{2}}{2} \log \left( \frac{b}{b_{c}}-1 \right) +\sqrt{2} \log 2 +O((b-b_{c})\log (b-b_{c})).
\end{equation} 
The regular part $I_{R}$ is defined by
\begin{equation}
I_{R}(b)\equiv \int^{1}_{0} (f(z,b)-f_{0}(z,b)) dz. 
\end{equation}
In the strong deflection limit $b\rightarrow b_{c}$, the regular part is given by
\begin{equation}\label{eq:IR}
I_{R}(b)=\sqrt{2}\log[2(2-\sqrt{2})] +O((b-b_{c})\log(b-b_{c})).
\end{equation}
From Eqs.~(\ref{eq:da}), (\ref{eq:I}), (\ref{eq:ID}), and (\ref{eq:IR}), 
the deflection angle in the strong deflection limit is obtained as
\begin{eqnarray}\label{eq:defstr_MWH}
\alpha(b)
&=&-\sqrt{2} \log \left( \frac{b}{b_{c}}-1 \right) +2\sqrt{2}\log[4(2-\sqrt{2})]-\pi \nonumber\\
&&+O((b-b_{c})\log (b-b_{c})).
\end{eqnarray}
Thus, we obtain $\bar{a}=\sqrt{2}$ and $\bar{b}=2\sqrt{2}\log[4(2-\sqrt{2})]-\pi$.

\subsubsection{The Schwarzschild spacetime}
We review a deflection angle in the Schwarzschild spacetime.
The line element is given by
\begin{equation}
ds^{2}=-\left( 1-\frac{2M_{s}}{\rho} \right)dt^{2}+\frac{d\rho^{2}}{1-\frac{2M_{s}}{\rho}} +\rho^{2}(d\theta^{2}+\sin^{2}\theta d\phi^{2}),
\end{equation}
where $M_{s}$ is a positive mass.
A light sphere exists at $\rho=3M_{s}$.
A light ray is scattered by a black hole if $b>b_{c}\equiv 3\sqrt{3}M_{s}$.
The deflection angle under the weak-field approximation $b \gg M_{s}$ is given by
\begin{equation}
\alpha(b)
=\frac{4M_{s}}{b}+ O\left( \left( \frac{M_{s}}{b} \right)^{2} \right).
\end{equation}
In the strong deflection limit $b\rightarrow b_{c}=3\sqrt{3}M_{s}$, the deflection angle becomes~\cite{Bozza_2002}
\begin{eqnarray}\label{eq:defstr_BH}
\alpha(b)
&=&-\log \left( \frac{b}{b_{c}}-1 \right) +\log[216(7-\sqrt{3})]-\pi \nonumber\\
&&+O((b-b_{c})\log (b-b_{c})).
\end{eqnarray}
Thus, we obtain $\bar{a}=1$ and $\bar{b}=\log[216(7-\sqrt{3})]-\pi$.

\subsubsection{The Ellis wormhole spacetime}
We briefly review the deflection angle of a light ray in an Ellis wormhole spacetime.
If $b>b_{c}\equiv a$, the light ray does not pass the wormhole throat 
and its deflection angle is given by~\cite{Chetouani_Clement_1984}
\begin{equation}
\alpha(b)=2K\left(\frac{a}{b}\right)-\pi. 
\end{equation}
Under the weak-field approximation $b \gg a$, the deflection angle becomes 
\begin{equation}
\alpha(b)=\frac{\pi}{4} \left(\frac{a}{b}\right)^{2} +O\left( \left( \frac{a}{b} \right)^{4} \right). 
\end{equation}
The deflection angle in the strong deflection limit $b\rightarrow b_{c}=a$ is obtained as~\cite{Tsukamoto:2016qro}
\begin{eqnarray}\label{eq:defstr_E}
\alpha(b)
&=&-\log \left( \frac{b}{b_{c}}-1 \right) +3\log 2 -\pi \nonumber\\
&&+O((b-b_{c})\log (b-b_{c})).
\end{eqnarray}
Thus, we get $\bar{a}=1$ and $\bar{b}=3\log 2 -\pi$.
We summarize $\bar{a}$, $\bar{b}$, and $b_{c}$ in Table~I.
\begin{table}[hbtp]
 \caption{$\bar{a}$, $\bar{b}$, and $b_{c}$ in the Schwarzschild, Ellis wormhole, and ultrastatic Schwarzschild-like (US) wormhole spacetimes.}
\begin{center}
\begin{tabular}{c c c c} \hline
                                        &$\bar{a}$          &$\bar{b}$                          &$b_{c}$  \\ \hline
Schwarzschild lens                      &$1$                &$\log[216(7-\sqrt{3})]-\pi$        &$3\sqrt{3}M_{s}$ \\ \hline
Ellis wormhole                          &$1$                &$3\log 2 -\pi$                     &$a$ \\ \hline
US wormhole                             &$\sqrt{2}$         &$2\sqrt{2}\log[4(2-\sqrt{2})]-\pi$ &$2M_{w}$ \\ \hline
\end{tabular}
\end{center}
\end{table}

\subsection{Microlens in a usual lens configuration under the weak-field approximation}
\begin{figure}[htbp]
\begin{center}
\includegraphics[width=70mm]{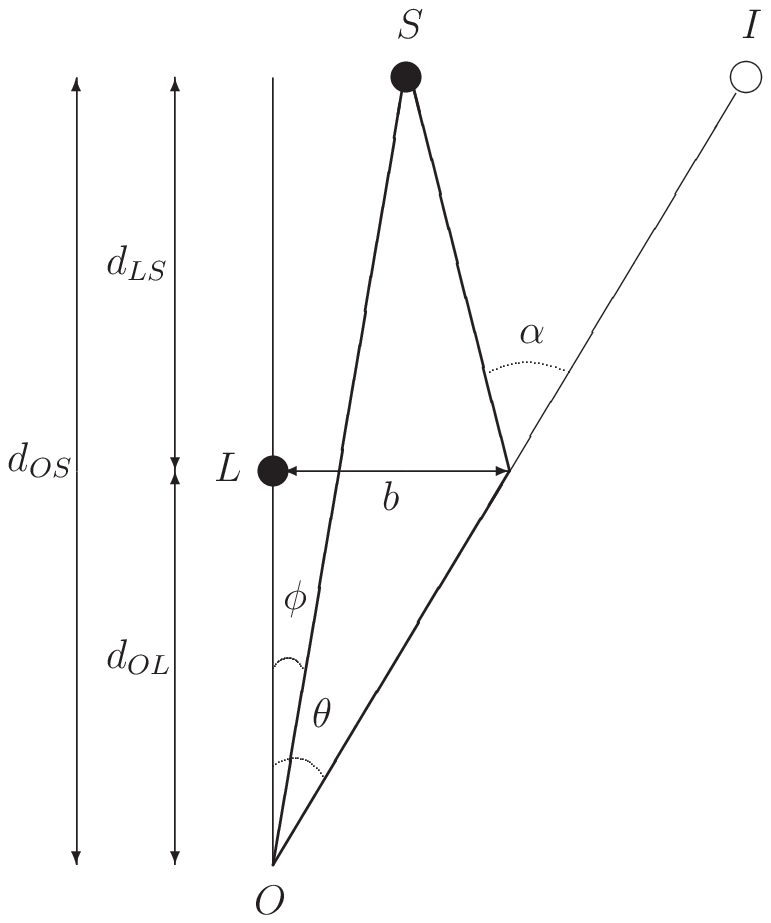}
\end{center}
\caption{A usual configuration for microlenses. A light ray emitted by a source at a source angle $\phi$ bends with a deflection angle $\alpha$ and 
reaches an observer with an image angle $\theta$. 
We assume $\left| \alpha \right| \ll 1$, $\left| \theta \right| \ll 1$, and $\left| \phi \right| \ll 1$.}
\label{Fig8}
\end{figure}

We consider microlenses~\cite{Paczynski_1986} in a usual lens configuration under the weak-field approximation.
Figure~\ref{Fig8} shows the lens configuration.
We consider a small angle lens equation given by
\begin{equation}\label{eq:lensweak}
d_{LS}\alpha=d_{OS}(\theta-\phi),
\end{equation}
where $\phi$ is a source angle, $\theta$ is an image angle, and $d_{LS}$ and $d_{OS}$ are the distances from a source to a lens and to an observer, respectively.
In this subsection, we assume that $\left| \alpha \right| \ll 1$, $\left| \theta \right| \ll 1$, and $\left| \phi \right| \ll 1$. 
The distance from the observer to the lens is given by $d_{OL}=d_{OS}-d_{LS}$.
Note that the impact parameter $b=d_{OL}\theta$ cannot be only positive but also negative in this subsection.

We consider a general spherical lens model with a deflection angle
\begin{equation}\label{eq:defweak}
\alpha = \pm C b^{-n} = \pm \frac{C}{d_{OL}^{n}\theta^{n}},
\end{equation}
where $C$ is a positive constant and $n$ is a positive integer~\cite{Tsukamoto_Harada_2013}.
The sign is a lower one if $\theta$ is negative and $n$ is even; otherwise it is upper one.
Under the weak-field approximation, 
the deflection angle is equivalent to the one in the Ellis wormhole spacetime when $n=2$ and $C=\pi a^{2}/4$,
the Schwarzschild spacetime when $n=1$ and $C=4M_{s}$,
and the ultrastatic Schwarzschild-like wormhole spacetime when $n=1$ and $C=2M_{w}$.

From Eqs.~(\ref{eq:lensweak}) and (\ref{eq:defweak}), the lens equation is rewritten as
\begin{equation}\label{eq:lensweak2}
\pm \hat{\theta}^{-n}=\hat{\theta}-\hat{\phi},
\end{equation}
where $\hat{\theta}\equiv \theta/\theta_{0}$ and $\hat{\phi}\equiv \phi/\theta_{0}$ 
and 
\begin{equation}\label{eq:Einstein}
\theta_{0}\equiv \left(  \frac{d_{LS}C}{d_{OS}d_{OL}^{n}} \right)^{\frac{1}{n+1}}
\end{equation}
is the unique positive solution of the lens equation for $\phi=0$, i.e., the Einstein ring angle.
Given $\phi$ and $n$, we find only a positive solution $\hat{\theta}_{+}$ and a negative solution $\hat{\theta}_{-}$~\cite{Tsukamoto_Harada_2013}. 
The total magnification $\mu_{tot}$ of the two images is given by
\begin{equation}
\mu_{tot}\equiv \left| \mu_{+} \right| + \left| \mu_{-} \right|,
\end{equation}
where the magnifications $\mu_{+}$ and $\mu_{-}$ of the positive image $\hat{\theta}_{+}$ and the negative image $\hat{\theta}_{-}$ are defined as
\begin{equation}
\mu_{+} \equiv \frac{\hat{\theta}_{+}}{\hat{\phi}}\frac{d\hat{\theta}_{+}}{d\hat{\phi}}
\end{equation}
and 
\begin{equation}
\mu_{-} \equiv \frac{\hat{\theta}_{-}}{\hat{\phi}}\frac{d\hat{\theta}_{-}}{d\hat{\phi}},
\end{equation}
respectively.

\begin{figure}[htbp]
\begin{center}
\includegraphics[width=80mm]{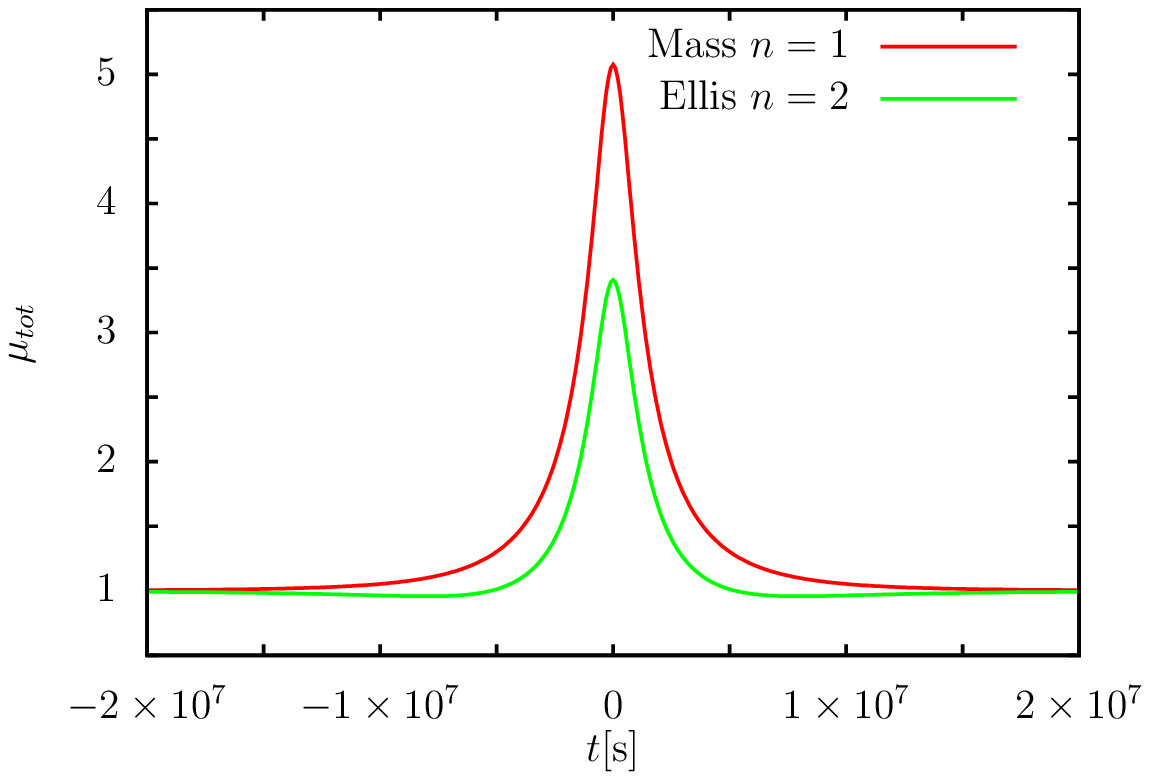}\\
\includegraphics[width=80mm]{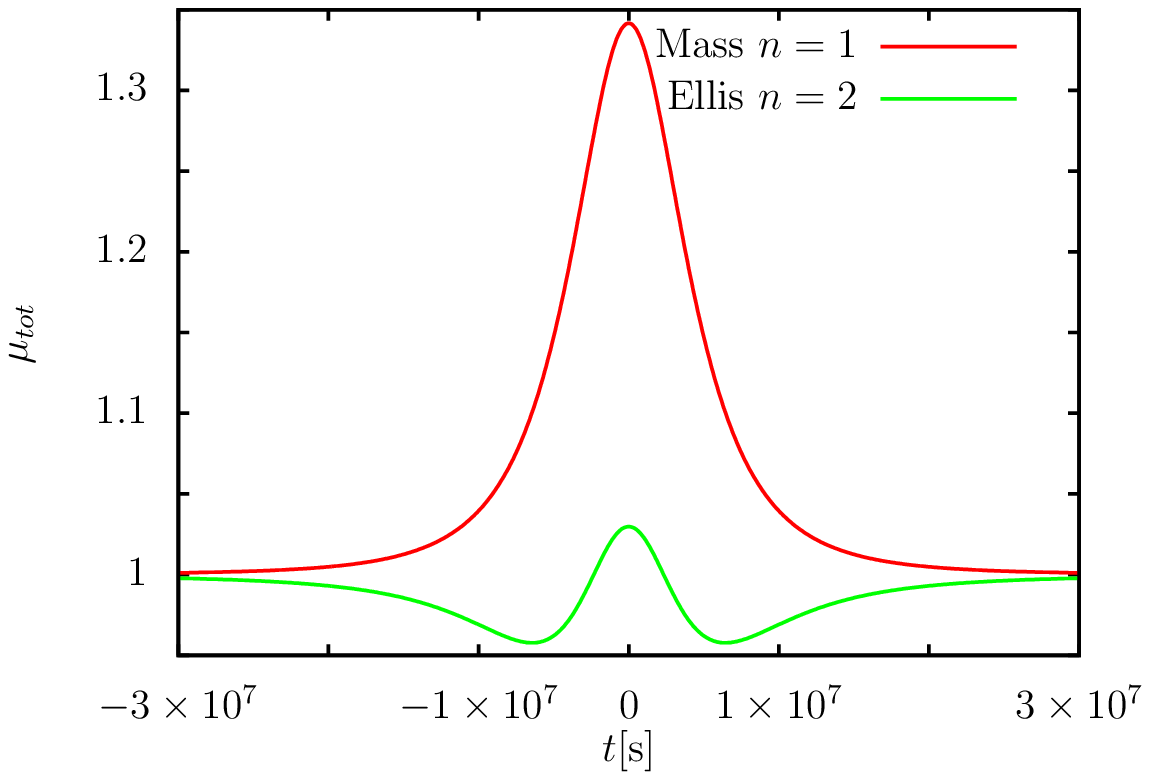}\\
\includegraphics[width=80mm]{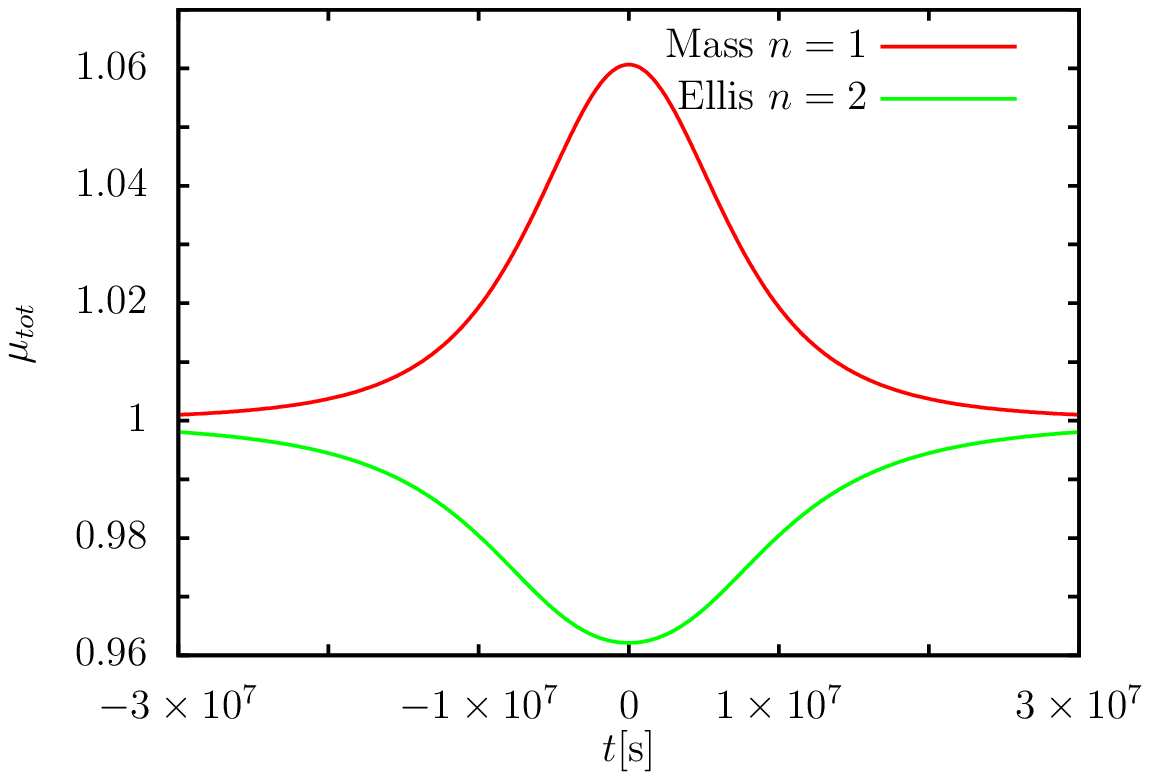}
\end{center}
\caption{Light curves of microlens.
The upper (red) and lower (green) curves denote light curves by a mass lens ($n=1$) 
with $M_{s}=1.5$km for a Schwarzschild lens or with $M_{w}=2M_{s}$ for an ultrastatic Schwarzschild-like wormhole 
and by an Ellis wormhole ($n=2$) with $a=4(2/\pi)^{1/2}(d_{LS}d_{OL}/d_{OS})^{1/4}M_{s}^{3/4}$, respectively.
They have the equal angle of the Einstein ring~$\theta_{0}$.
The light curves of the two mass lenses are the same under the weak-field approximation.
We assume that $d_{OL}=d_{LS}=10$kpc and $d_{OS}=20$kpc and a source moves with the velocity $200$km/s on the source plane. 
The top, middle, and bottom panels show the light curves with the closest separations $0.2\theta_{0}$, $\theta_{0}$, and $2\theta_{0}$, respectively, 
between the optical axis $\phi=\pi$ and the position of the source.}
\label{Fig9}
\end{figure}

We consider the light curves of microlenses by massive objects and an Ellis wormhole.
We assume $d_{OL}=d_{LS}=10$kpc and $d_{OS}=20$kpc and a source moves with the velocity $200$km/s on the source plane. 
We set $M_{s}=1.5$km, $M_{w}=2M_{s}=3$km, and
$a=4(2/\pi)^{1/2}(d_{LS}d_{OL}/d_{OS})^{1/4}M_{s}^{3/4}=8.6\times
10^{4}$km so that the three lenses have the same value for the 
Einstein ring $\theta_{0}$ according to Eq.~(\ref{eq:Einstein}).
Note that the light curves caused by the Schwarzschild lens and by the ultrastatic Schwarzschild-like wormhole are the same if $M_{w}=2M_{s}$.
The light curves are shown in Fig.~\ref{Fig9}.
We see that the light curves in the Ellis wormhole spacetime have demagnified periods as pointed out by Abe~\cite{Abe_2010}.
The demagnified light curves apparently violate a known magnification theorem that the total magnification by an isolated mass lens is always larger than unity.
The Ellis wormhole, however, is not described by an isolate mass lens since it has a vanishing ADM mass and its gravitational potential is asymptotically proportional to $1/r^{2}$. 

In the rest of this subsection, we show the existence of the demagnification by an Ellis wormhole in a simple analytical calculation.
We concentrate on the behaviors of the magnification at $1 \ll \hat{\phi} \ll \theta_{0}^{-1}$ assuming $\theta_{0} \ll 1$.
The angle and the magnification of images are obtained as
\begin{equation}
\hat{\theta}_{+}(\hat{\phi})\sim \hat{\phi}+\hat{\phi}^{-n}-n\hat{\phi}^{-2n-1}
\end{equation}
\begin{equation}
\hat{\theta}_{-}(\hat{\phi})\sim -\hat{\phi}^{-\frac{1}{n}}
\end{equation}
and
\begin{equation}\label{eq:mu+}
\mu_{+}(\hat{\phi})\sim 1-(n-1)\hat{\phi}^{-n-1}+n(2n-1)\hat{\phi}^{-2n-2}
\end{equation}
\begin{equation}
\mu_{-}(\hat{\phi})\sim  -\frac{1}{n}\hat{\phi}^{-\frac{2}{n}-2},
\end{equation}
respectively. 
Notice that the subleading terms depend on $n$ and that the second term of the right-hand side in Eq.~(\ref{eq:mu+}) vanishes when $n=1$. 
The total magnifications are $\mu_{tot}\sim 1+2\hat{\phi}^{-4}$, $\mu_{tot}\sim 1-(1/2)\hat{\phi}^{-3}$, and $\mu_{tot}\sim1+(1/n)\hat{\phi}^{-2/n-2}$ 
for $n=1$ (mass lens),  $n=2$ (Ellis wormhole), and $n>2$, respectively.  

The derivatives of $\left| \mu_{+} \right|$ and $\left| \mu_{-} \right|$ with respect to $\hat{\phi}$ are given by
\begin{equation}
\left| \mu_{+} \right|'\sim (n-1)(n+1)\hat{\phi}^{-n-2}-2n(2n-1)(n+1)\hat{\phi}^{-2n-3}
\end{equation}
and
\begin{equation}
\left| \mu_{-} \right|'\sim -\frac{2}{n} \left( \frac{1}{n}+1 \right) \hat{\phi}^{-\frac{2}{n}-3},
\end{equation}
respectively, where $'$ denotes differentiation with respect to $\hat{\phi}$.
For $n=1$ (mass lens), $n=2$ (Ellis wormhole), and $n>2$, the derivatives of the total magnification are obtained as 
$\mu_{tot}'\sim -8\hat{\phi}^{-5}$, $\mu_{tot}'\sim (3/2)\hat{\phi}^{-4}$, and $\mu_{tot}'\sim-2n^{-2}(1+n)\hat{\phi}^{-2/n-3}$, respectively.
Hence, the light curves for $n=1$ (mass lens) have at least one local maximum that is larger than unity
while the ones for $n=2$ (Ellis wormhole) have at least one local minimum that is smaller than unity.~\footnote{
Kitamura \textit{et al.} showed numerically that light curves are demagnified when $n>1$~\cite{Kitamura_Nakajima_Asada_2013}.
Our analytical approach, however, seems to be unsuitable for showing the existence of the demagnification in the $n>2$ case.}

Figure~\ref{Fig9} shows that
$\mu_{tot}$ monotonically increases
in time, reaches the maximum peak value, which is greater than unity, 
and then monotonically decreases 
for $n=1$. However, for $n=2$ (the Ellis wormhole), $\mu_{tot}$ first 
decreases. The following behavior is divided into two cases, depending 
on 
the value of the closest separation.
 In the first case, it monotonically decreases
to the unique minimum value, which is smaller than unity, and then
monotonically increases. In the second case, it decreases to the first minimum, which is smaller
than unity, increases to a local maximum value, decreases 
to the second minimum, which is the same as the first one, 
and then monotonically increases.

\subsection{Retrolensing}
\begin{figure}[htbp]
\begin{center}
\includegraphics[width=70mm]{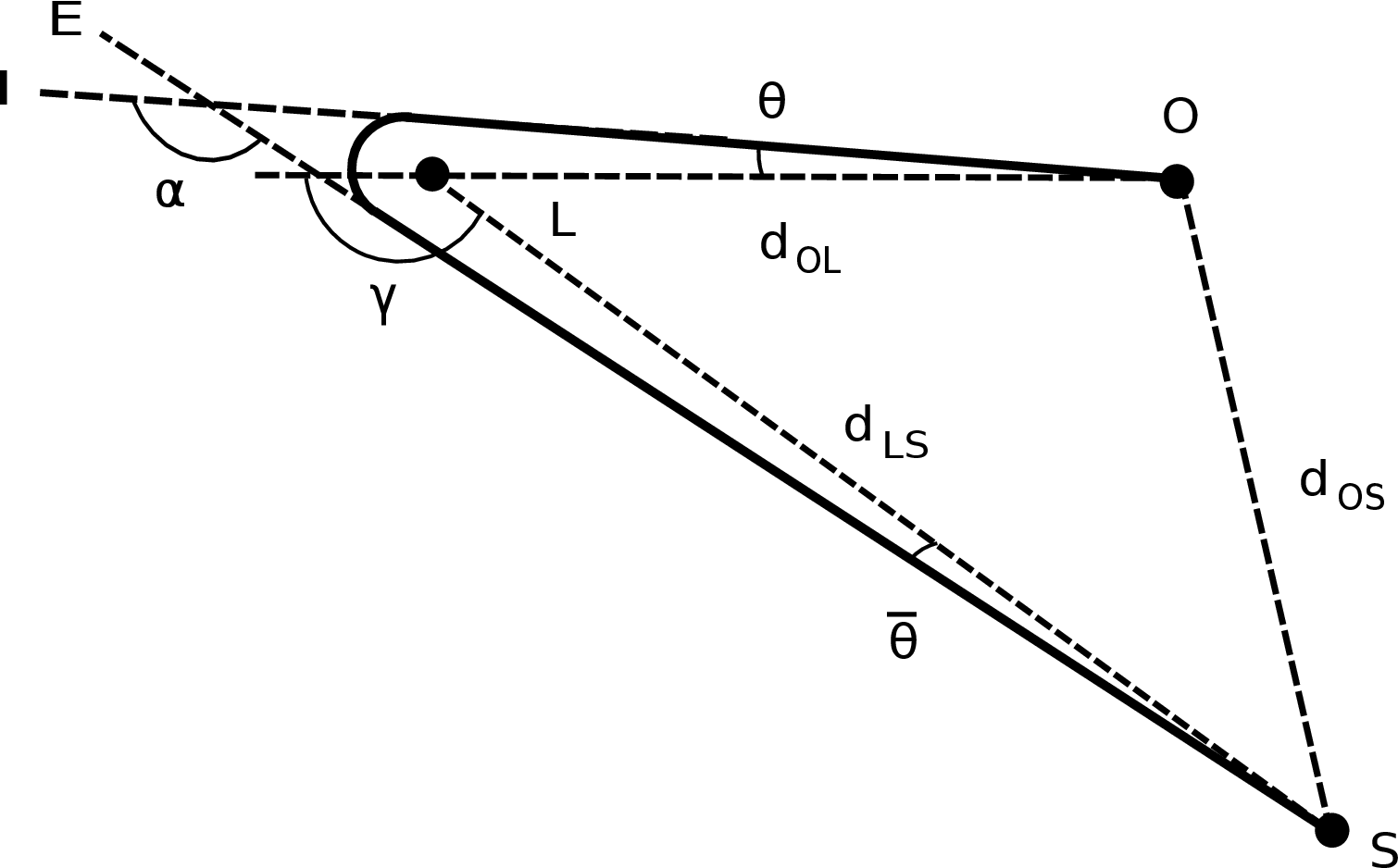}
\end{center}
\caption{ Configuration of a retrolens. A light ray emitted in a direction $E$ by a source $S$ with a source angle $\gamma\sim \pi$ 
bends near a lens $L$ with the deflection angle $\alpha$ and reaches an observer $O$ with an image angle $\theta$.
$\bar{\theta}$ is an angle defined by $\angle ESL$.}
\label{Retro}
\end{figure}

In this subsection, we discuss light curves of retrolensing~\cite{Holz:2002uf,Eiroa:2003jf,Bozza:2004kq}.
We concentrate on a case where a point source $S$ emits a light ray in a direction $E$ 
and it is reflected near a light sphere of a lens $L$. An observer $O$ sees an image $I$.
The lens configuration is shown in Fig.~\ref{Retro}.
We use a lens equation proposed by Ohanian~\cite{Ohanian_1987} and discussed in~\cite{Bozza:2004kq,Bozza:2008ev},
\begin{equation}\label{eq:Ohanian}
\gamma=\alpha-\theta-\bar{\theta},
\end{equation}
where
$\gamma$ is a source angle defined as the supplementary angle of $\angle OLS$,
$\alpha$ is the deflection angle, 
$\theta$ is an image angle given by $\angle IOL$,
and $\bar{\theta}$ is $\angle ESL$.
We concentrate on a positive impact parameter $b$.
We assume that the lens, the observer, and the source are almost aligned in this order.
From this assumption, we obtain $\gamma \sim \pi$ and $d_{LS}=d_{OL}+d_{OS}$.
We also assume that the lens and the source are far away from the observer, i.e,
$b\ll d_{OL}$ and  $b\ll d_{OS}$,
and we neglect both $\theta=b/d_{OL}$ and $\bar{\theta}=b/d_{LS}$ in the Ohanian lens equation.

From Eqs.~(\ref{eq:defstr}) and (\ref{eq:Ohanian}), the positive solution of the Ohanian lens equation is given by
\begin{equation}
\theta=
\theta_{+}
\equiv
 \theta_{c} \left[ 1+\exp\left( \frac{\bar{b}-\gamma}{\bar{a}} \right) \right],
\end{equation}
where $\bar{a}$ and $\bar{b}$ are constant numbers in the deflection angle~(\ref{eq:defstr}) 
in the strong deflection limit $b\rightarrow b_{c}$
and where $\theta_{c}\equiv b_{c}/d_{OL}$ is the image angle of a light sphere.
A negative solution $\theta_{-}$ is given by $\theta=\theta_{-}=-\theta_{+}$.
The total magnification of the two images is obtained as~\cite{Bozza:2004kq}
\begin{equation}
\mu_{tot}(\gamma)\sim 2 \left(\frac{d_{OS}\theta_{c}}{d_{LS}}\right)^{2}
\frac{e^{(\bar{b}-\gamma)/\bar{a}}\left[ 1+e^{(\bar{b}-\gamma)/\bar{a}} \right]}{\bar{a}\sin\gamma}.
\end{equation}
See Table~I for $\bar{a}$, $\bar{b}$, and $b_{c}=\theta_{c}d_{OL}$.
We assume $d_{OL}=d_{OS}=10$kpc and $d_{LS}=20$kpc and the source moves with the velocity $200$km/s on the source plane. 
We set $M_{s}=1.5$km, $M_{w}=9\sqrt{3}/4$km, and $a=9\sqrt{3}/2$km so that the image angles of their light spheres are the same: 
$\theta_{c}\sim 2.5 \times 10^{-17}$ rad.
Figure~12 shows the light curves of the retrolenses.
We notice that the shapes of the light curves of the retrolenses look like 
the ones of light curves formed by light rays passing an Ellis wormhole throat in Figs.~\ref{Fig6} and \ref{Fig7}. 
\begin{figure}[htbp]
\begin{center}
\includegraphics[width=80mm]{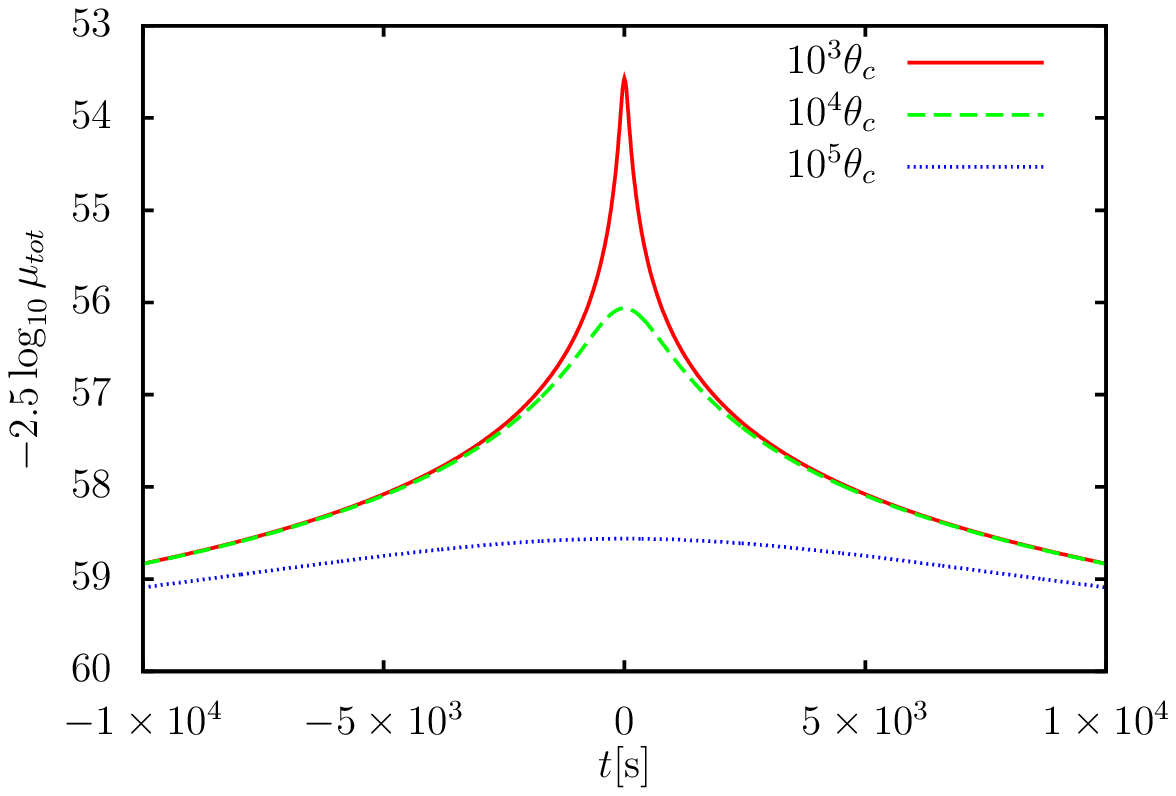}\\
\includegraphics[width=80mm]{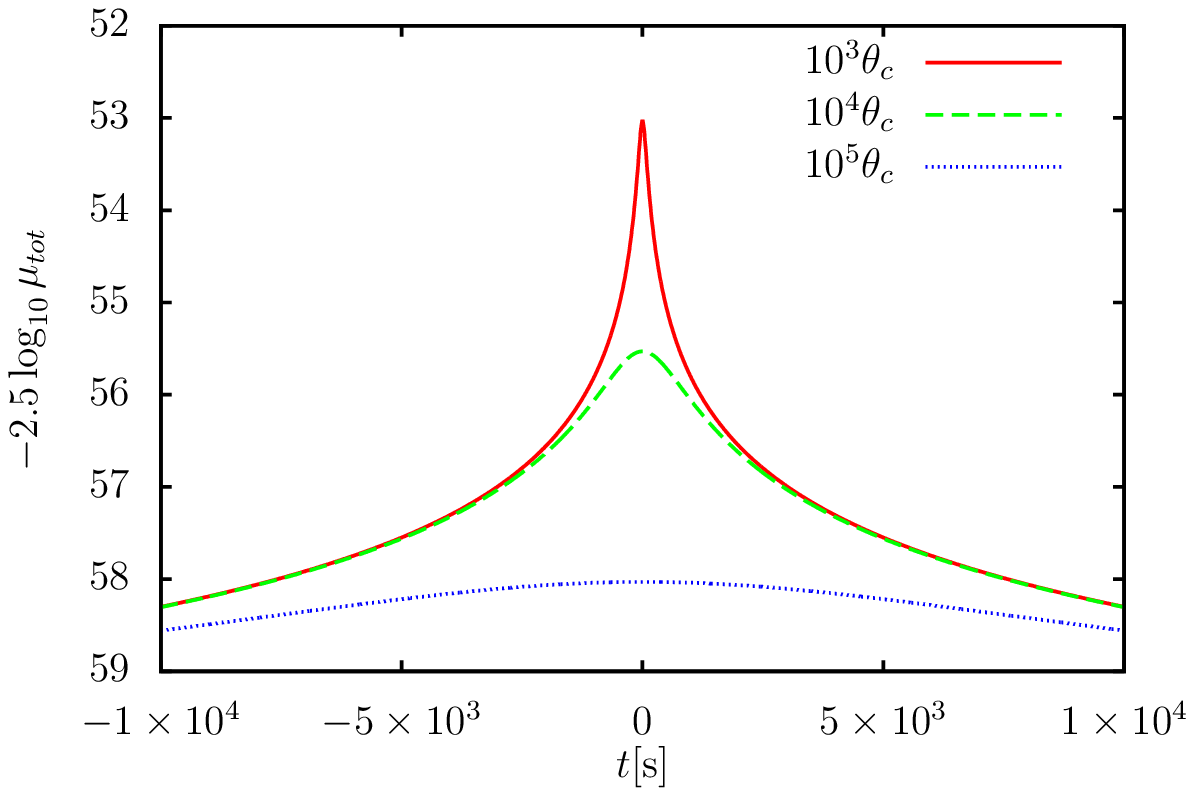}\\
\includegraphics[width=80mm]{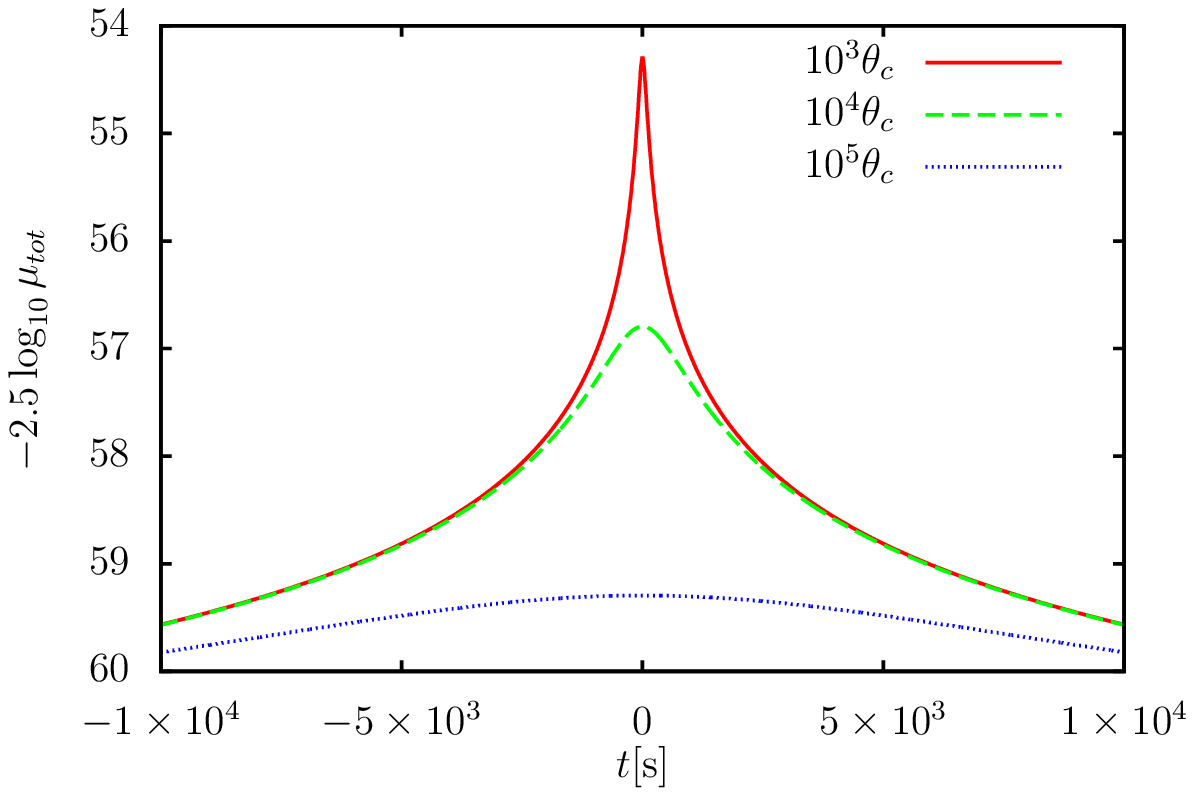}
\end{center}
\caption{Light curves of retrolensing.
The top, middle, and bottom figures are light curves in the Schwarzschild spacetime with $M_{s}=1.5$km, in the massive wormhole spacetime with $M_{w}=9\sqrt{3}/4$km, 
and in the Ellis wormhole spacetime with $a=9\sqrt{3}/2$km, respectively.
The parameters of the spacetimes are tuned so that they have the equal size of the image angles $\theta_{c}$ of the light spheres.
We assume $d_{OL}=d_{OS}=10$kpc and $d_{LS}=20$kpc and a source moves with the velocity $200$km/s on the source plane. 
The solid (red), broken (green), and dotted (blue) curves denote light curves 
with the closest separations $10^{3}\theta_{c}$, $10^{4}\theta_{c}$, and $10^{5}\theta_{c}$, respectively, between the optical axis $\gamma=\pi$ and the source position. }
\label{Fig11}
\end{figure}

\section{Concluding Remarks}
We have investigated light curves
 due to light rays passing through an Ellis wormhole throat with the throat radius $a$.
We concentrate on the cases where a source passes by $\phi = \pi$ and 
$\phi = 0$.

In the case where the source passes by $\phi=\pi$,
a pair of the primary and secondary images 
is the brightest among 
an infinite number of images.
Tertiary and higher-order images are fainter than them.
In the case where the source passes by $\phi=0$, 
a pair of the secondary and tertiary images 
has a dominant contribution to the light curve. 
The two images are a pair of relativistic images due to light rays 
that have passed around the light sphere with almost symmetric orbits.
The primary, quaternary, and higher-order images are fainter than them.
By comparing the two cases, we notice that the former pair is 
brighter than the latter
pair if the closest separations $\beta$ are the same.

In the both cases, 
the light curves of the pair of images are similar both in shape and magnitude.
The time difference of the two peaks of the light curves is given by $2a\beta$.
If the time difference is too short, an observer does not separate 
the two light curves and 
he or she regards them as a single light curve with a single peak. 

The closest separation $\beta$ between the source and the axis 
$\phi=0$ or $\phi=\pi$ 
determines the peak magnitude of the light curve. 
The observer can observe a highly amplified light curve
if $\beta$ is very small. 
The time scale of the 
light curves
 depends on the velocity 
of the source projected on the source plane. 

When the source passes by $\phi = \pi$, a pair of the tertiary and 
quaternary images 
appears slightly outside a pair of the primary and secondary images.
The light rays of the tertiary and quaternary images reach 
the observer later 
than the ones of the primary and secondary images.
Since $T-T_{0} \sim a \left| \Phi \right| +$const 
for $\left| \Phi \right| \gtrsim \pi$, 
the time difference between the two pairs is estimated to $2\pi a$,
which is much longer than $2a\beta$, 
the time difference of the peaks of the primary and 
secondary images.
If the time scale of the 
lensing is shorter than $2\pi a$, 
the observer can separate the second peak of the light curve 
from the first peak.
If we distinguish the two peaks of the two pairs, 
we can estimate the proper length of the throat 
from the observed time difference.
This is also the case for a source passing by $\phi = 0$.

We have investigated the light curves of retrolensing by an Ellis
wormhole, an ultrastatic Schwarzschild-like wormhole, and a Schwarzschild black hole.
We notice that the shapes of light curves of light rays passing through an Ellis wormhole throat look like 
the ones of light curves of the retrolenses.
From this fact, we make a conjecture that the shapes of the light curves of a point source made by light rays passing by light spheres 
do not depend on the details of static spherically symmetric and asymptotically flat spacetimes and gravitational lens configurations 
but the apparent magnification of observed light curves relies on them.
If the conjecture is true, we cannot distinguish the other Morris-Thorne wormholes including the Ellis wormhole and the ultrastatic Schwarzschild-like wormhole
from black holes with the shape of light curves related to light spheres. 

Since the Ellis wormhole has a vanishing ADM mass and its gravitational potential is asymptotically proportional to $1/r^{2}$,
a well-known magnification theorem that the total magnification of images lensed 
by an isolated mass is always larger than unity under the weak-field approximation
cannot be applied to gravitational lenses by the Ellis wormhole.
In fact, under the weak-field approximation,
the light curve of the sum of primary and secondary images in the Ellis
wormhole spacetime has gutters on both sides of the peak, 
if the source and the observer are on the same side of the throat or
$r_{S}r_{O}>0$~\cite{Abe_2010}.
In this paper, we have shown analytically the existence of demagnified light curves in the Ellis wormhole spacetime under the weak-field approximation.
Thus, if we observe both the characteristic demagnified light curves
and the characteristic light curves caused by their light sphere, which have been found in the current
paper, the lensing object can be regarded as a candidate of an Ellis wormhole.

Our method can also apply for traversable wormhole spacetimes with a positive mass. 
In the current paper, we have considered a light ray that is 
emitted by a point source and then strongly deflected by a
lens object in the vicinity of its light sphere. 
We conjecture that the light curves of such light rays are 
very similar in shape, whether the lens is an isolated mass or 
a wormhole. This conjecture should be tested in future work.
Given light curves of light rays deflected by light spheres, the absolute brightness of the source, 
and the details of the lens configuration such as distances and the closest separation $\beta$,
we obtain not only the mass but also information on the full metric of spacetimes. 
It is well known that Morris-Thorne wormholes violate the weak energy condition at least at the throat if we assume general relativity~\cite{Morris_Thorne_1988}.
The detection of light curves made by light rays deflected by a light sphere investigated in this paper 
does not tell immediately the existence of violation of the weak energy condition
but if we identify the details of the lens
configuration by wormholes and gravitational theory, 
the observed light curves provide evidence for violation of the weak energy condition.
We hope that this paper stimulates further work in this direction.

\section*{ACKNOWLEDGMENTS}
The authors thank U.~Miyamoto for letting N.~T. know the relevant work of Perlick~\cite{Perlick_2004_Phys_Rev_D} 
at the very early stage of this work. 
They also thank H. Asada, T. Kitamura, T. Igata, M. Patil, S. Yokoyama,
Y. Gong, T. Shiromizu, and T. Kobayashi for valuable comments and discussion.
They thank an anonymous referee for valuable comments and suggestions.
N.~T. acknowledges support for this work by the Natural Science Foundation of China under Grants No. 11475065
and the Program for New Century Excellent Talents in University under Grant No. NCET-12-0205.
T.~H. was supported by JSPS KAKENHI Grant No. JP26400282.
%
%
%

\end{document}